\documentclass[10pt]{article}
\usepackage{textcomp}
\usepackage{a4wide}
\usepackage{graphicx}
\usepackage{amsmath}
\usepackage{amsbsy}
\usepackage{epstopdf, epsfig}
\usepackage{amssymb}
\usepackage{longtable}
\usepackage{mathtools}
\usepackage{textcomp, csquotes}
\usepackage{xfrac} 
\usepackage[inline]{enumitem}
\usepackage{afterpage}
\usepackage{natbib}
\usepackage{subfigure}

\usepackage{gensymb}
\usepackage{setspace}
\usepackage{multirow}
\usepackage{geometry}
\usepackage{lineno}
\usepackage[hidelinks,colorlinks=true,linkcolor=blue]{hyperref}
\hypersetup{citecolor=blue, urlcolor=blue}
\usepackage{xspace}
\newcommand{\co}{CO\ensuremath{_\textrm{2}}\xspace}
\newcommand*{\doi}[1]{\href{http://dx.doi.org/#1}{doi: #1}}

\title{Field-scale Impacts of Long-Term Wettability Alteration in Geological \co Storage}

\author{A.M. Kassa$^\text{a,b}$ \and
  S.E. Gasda$^\text{*,a}$\and
  D. Landa-Marb\'an$^\text{a}$ \and
  T.H. Sandve$^\text{a}$\and
  K. Kumar$^\text{b}$ \and
  }
\date{}
\begin{document}
\pagenumbering{arabic}
\maketitle
\onehalfspace
\noindent ${}^\text{a}$ Department of Energy \& Technology, NORCE Norwegian Research Centre AS, Norway.\\[5pt]
${}^\text{b}$ Department of Mathematics, University of Bergen, Norway.\\[5pt]
${}^\text{*}$ Corresponding author: Sarah E. Gasda (E-mail address: sgas@norceresearch.no).
\vspace{.25cm}
\begin{abstract}
\noindent Constitutive functions that govern macroscale capillary pressure and relative permeability are central in constraining both storage efficiency and sealing properties of \co storage systems. Constitutive functions for porous systems are in part determined by wettability, which is a pore-scale phenomenon that influences macroscale displacement. While wettability of saline aquifers and caprocks are assumed to remain water-wet when \co is injected, there is recent evidence of contact angle change due to long-term \co exposure. Weakening of capillary forces alters the saturation functions dynamically over time. Recently, new dynamic models were developed for saturation functions that capture the impact of wettability alteration (WA) due to long-term \co exposure. In this paper, these functions are implemented into a two-phase two-component simulator to study long-term WA dynamics for field-scale \co storage. We simulate WA effects on horizontal migration patterns under injection and buoyancy-driven migration in the caprock. We characterize the behavior of each scenario for different flow regimes. Our results show the impact on storage efficiency can be described by the capillary number, while vertical leakage can be scaled by caprock sealing parameters. Scaling models for \co migration into the caprock show that long-term WA poses little risk to \co containment over relevant timescales.   
\end{abstract}

\begin{centering}
\includegraphics[scale=.6]{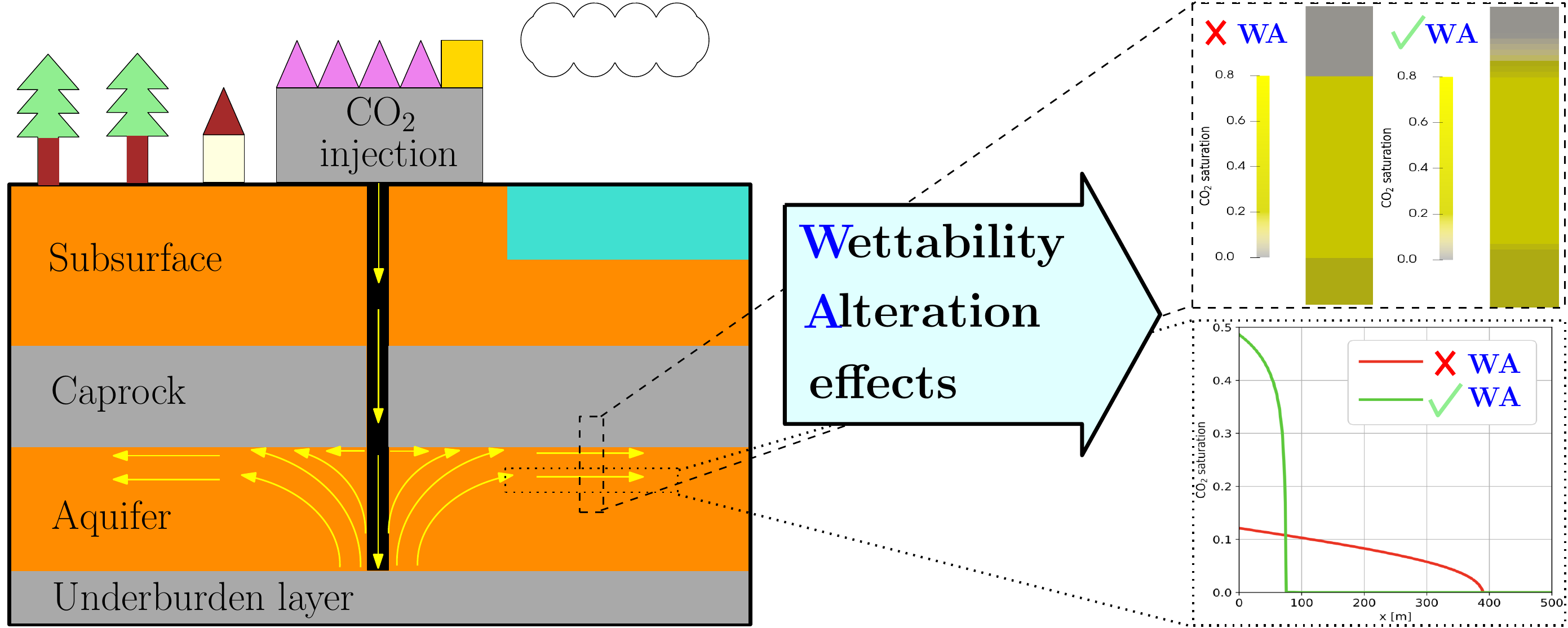}
\end{centering}


\thispagestyle{empty}

\section{Introduction}\label{intro} 
Geological \co storage can be successfully implemented in deep saline aquifers that possess favorable characteristics \citep{IPCC2005}, e.g., the storage formation has sufficient capacity, injectivity, and containment properties to store the desired quantity of \co at the required injection rate. A competent caprock should be verified that provides a capillary and permeability seal to prevent buoyant \co from migrating upwards and is free of defects that could leak or seep \co out of the storage formation \citep{Busch2008}. In addition, lateral migration of \co should stay within the defined boundaries of the storage reservoir. {In some cases, the boundaries are defined by a structural trap. In open stores with no defined trap, sufficient \co trapping efficiency in residual and dissolved phase is needed to prevent unwanted lateral \co migration \citep{Tucker2018}. } 

{There is a wide array of geological and operational factors affecting \co storage migration and containment in deep, saline aquifer \citep{Birkholzer2015}. The physical and chemical rock-fluid interactions play an important role through their impact on the constitutive functions that govern macroscale capillary pressure and relative permeability. These functions, often called flow functions or saturation functions, govern the displacement processes in a multiphase flow setting and are central in constraining both storage capacity of the formation and sealing properties of the caprock \citep{Mori2015, Negara2011, Oostrom2016, Sarkarfarshi2014}.} Capillary pressure and relative permeability functions for the storage reservoir and caprock are strongly influenced by wettability. The term wettability refers to the preference of one fluid over the other for the rock surface \citep{Bonn09, Eral12} and is a pore-scale phenomenon that has direct influence on macroscale displacement processes \citep{Anderson1987, Bobek1958, Olugbenga2014}.

{Wettability has been shown to play a key role in geological \co storage with regard to \co migration, trapping, and sealing capacity.} 
Saline aquifers are brine-bearing formations, and thus often assumed to be strongly water-wet. 
{However, differences in initial wetting state can be caused by a number of factors that affect the surface chemistry of rock minerals, including temperature and pressure conditions, brine salinity, mineralogy, and organic content (see \cite{Iglauer2014} and references therein). Wettability significantly affects the capillarity between \co and brine, which in turn impacts the efficiency of brine displacement. Studies have shown that wettability affects plume migration \citep{Al-Khdheeawi2017} and residual trapping \citep{Krevor2015}.} In terms of containment, the capillary seal of the caprock barrier is reliant on a strongly water-wet condition to maintain a high capillary force in combination with smaller pore radii of a tight medium. An effective seal increases storage capacity in structural traps by supporting a thick column of \co \citep{Iglauer2014}, which can be compromised if the seal weakens over time due to changes in wettability \citep{Rezaee2017}.

{Although wettability is often considered a static property, there is a possibility for wettability to change dynamically over time.} This process of \textit{wettability alteration} (WA) is a dynamic process, that involves a complex interaction of surface mineralogy, fluid composition, and reservoir conditions that is percipitated by the introduction of a wettability-altering agent \citep{Bonn09, Buckley1998}. WA has been long studied in the petroleum sector, and a variety of methods and agents, that include solvents such as supercritical \co, have been applied to enhance WA in oil reservoirs to the benefits of increased recovery \citep{Drexler2020, Sun2017}. Modeling of WA often assumes an instantaneous change of wettability as a function of solvent concentration \citep{RefLashgari}.  However, there are some cases where a WA process depends on exposure time that can extend for weeks or months to the wettability-altering agent \citep{Blunt2017, Farokhpoor2013, Fatah2021, Gholami2021, Jafri2016, Saraji2013}. 

For \co storage, there has been some intriguing evidence of long-term WA caused by \co exposure that significantly weakens the strength of capillary forces of a water-wet medium, which can apply to both reservoir rocks and caprock material. For example, core-flooding experiments and capillary pressure measurements on storage reservoir samples have reported a instability or slow reduction in capillary pressure--saturation relationships over time \citep{Farokhpoor2013, Plug07, Tokunaga2013, RefW, Wang13, RefWang}. In particular, we note in \cite{Wang13} that wettability was measured through contact angle (CA) change in silicate and carbonate sands from strongly water wet to intermediate wet over the course of weeks and months.  Studies on shales exposed to \co over long periods show a steady increase in shale/water CA over days and months \citep{Chiquet07a, Fatah2021, Gholami2021}, with trends that indicate additional exposure  would lead to further CA increase. {Similar studies using reservoir rocks also show an evolution from water- to \co-wet conditions due \co exposure in core-flooding experiments \citep{Fauziah2021, Valle2018}.} All these bench-scale experiments indicate that WA introduces long-term dynamics into the capillary pressure and relative permeability functions. Such dynamics could have a significant impact on \co storage at the field scale in ways that may be positive or negative. On the one hand, a caprock that becomes intermediate-wet or hydrophobic is less effective than a water-wet caprock in providing an effective capillary barrier.  On the other hand, capillary diffusion may be reduced in the storage reservoir through long-term WA, which can be beneficial to injectivity and capacity by causing the \co to displace the resident fluid efficiently \citep{Kassa2020b}. A wettability change, in general, affects the space-time evolution of \co plume in the formation in ways that has not been fully understood yet. 

{There have been several attempts to understand the impact of static wettability on long-term \co storage at field scales. In \cite{Al-Khdheeawi2017, Al-Khdheeawi2018, Al-Khdheeawi2017b} the main conclusions are that \co vertical migration is enhanced with more \co-wet systems, which is due to the lower residual trapping capacity of \co-wet rocks. The impact is stronger when considering heterogeneous wettability and increased temperature. These studies did not investigate the impact of \co exposure that changes rock wettability over time. Another study investigated wettability alteration caused by increased reservoir temperature under \co injection. The relative permeability curves were set in predetermined zones different temperature, but the curves were not adjusted dynamically in time \citep{Abbaszadeh2020}.}

{Understanding how \co storage is impacted by dynamic wettability caused by \co exposure requires a different approach than previously used to study static wettability. Two simulation components are needed:} (a) flow functions that capture the dynamics due to \co exposure, and (b) analysis and characterization of long-term WA impacts through implementation of dynamic flow functions into field-scale simulation. We have recently addressed the first aspect in \cite{Kassa2019} and \cite{Kassa2020} where we successfully developed dynamic capillary pressure and relative permeability functions through mechanistic modeling combined with upscaling of wettability dynamics from pore to core scale. The developed models are extended forms of standard constitutive models either by interpolation between two end-state capillary pressure functions \citep{Kassa2019} or by incorporating WA dynamics directly into the parameters of the existing relative permeability model \citep{Kassa2020}. The dynamics in both models are driven by exposure time to \co which is easily calculated from the local saturation history. We found that both models compare well against pore-scale simulation data where WA is modeled in individual pores. More interestingly, we determined that the upscaled model parameters have a clear relationship with the underlying pore-scale WA model parameter. Moreover, the developed constitutive models are robust and lend themselves to implementation within standard flow simulators.

{To our knowledge, numerical simulation of the field-scale impacts of WA caused by long-term \co exposure has not been performed previously.} In this paper, we implement the aforementioned dynamic relative permeability and capillary pressure functions  \citep{Kassa2019, Kassa2020} into flow models for \co storage applications in the open porous media (OPM) framework \citep{Rasmussen2019}. We investigate the impact of long-term WA process for \co storage with a focus on the migration patterns and containment in a saline aquifer that involves both viscous-driven flow in the storage aquifer and the potential for buoyancy flow through the caprock. Therefore, we concentrate our analysis on one-dimensional horizontal and vertical flow systems in order to understand the impact of {dynamic changes in wettability} on both flow regimes separately. For the horizontal case, we study the impact of WA on storage efficiency given by the front location, while for the vertical case, we focus on the impact of \co containment under a capillary seal (i.e., caprock). In both cases, we study the sensitivity of {wettability dynamics and subsequent field-scale impacts} under different conditions (i.e., permeability, entry pressure, injection rate, and porosity) and ultimately seek an appropriate scaling relationship for the quantity of interest in order to generalize our findings. The quantified behavior contributes to increased understanding of the efficiency and integrity of long-term \co storage subject to WA {caused by \co exposure}.

\section{Methodology}\label{NonlocalTPF}

\begin{figure}[t!]
\centering
\includegraphics[scale=0.35]{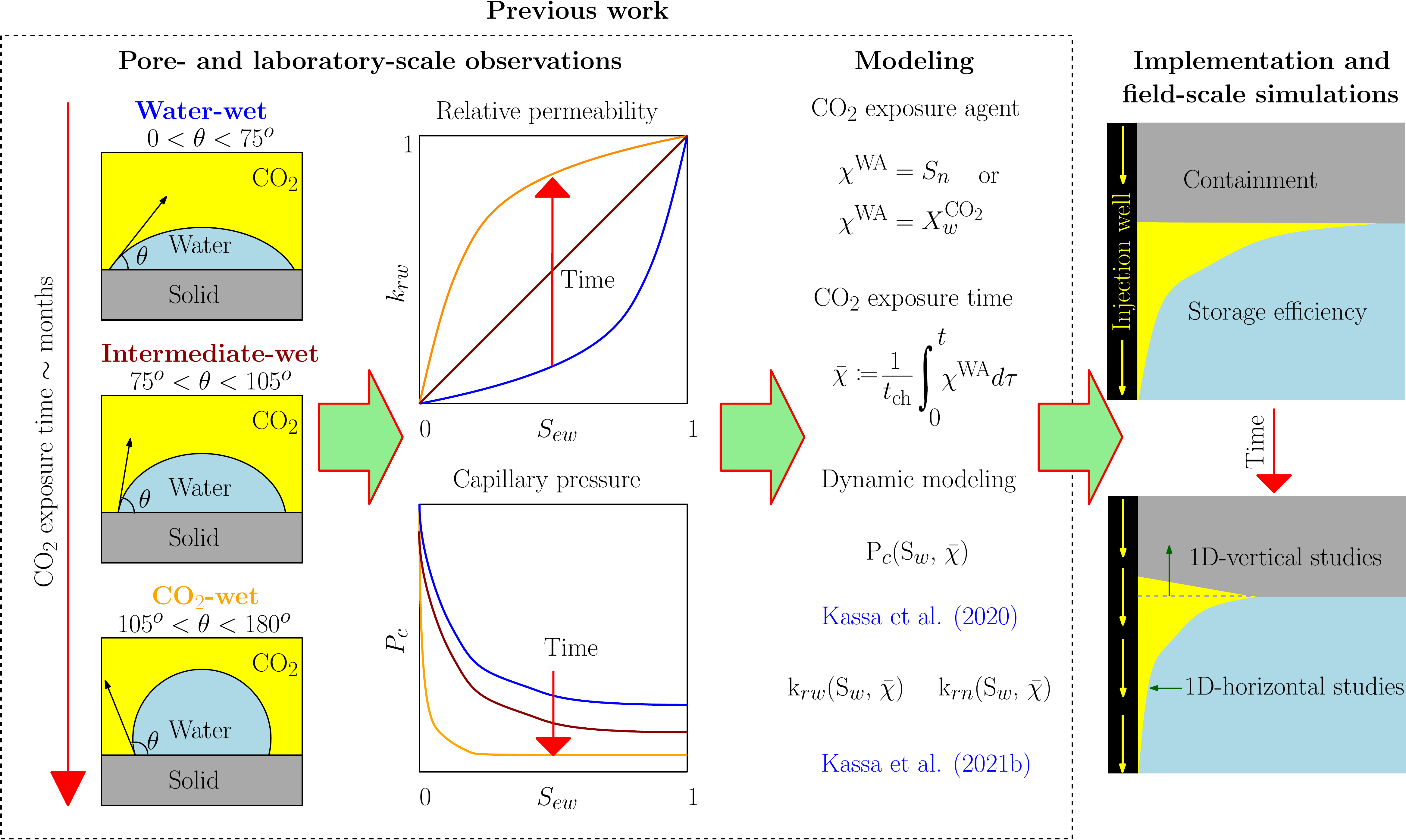}
\caption{{Model development involved in preparation for long-term wettability alteration field-scale studies. The boxed content illustrates key features of previous work \citep{Kassa2019, Kassa2020}  implemented in this study. Further discussion of the ``Modeling'' formulations can found in Section \ref{SatuFunc}. }}\label{fig:data} 
\end{figure}

{The goal of this study is to investigate the impacts of dynamic changes in wettability caused by long-term \co exposure on field-scale \co plume evolution and containment. In this section, we outline the methods that have been implemented to carry out the desired field-scale simulations. Figure \ref{fig:data} shows the relevant components of the dynamic saturation functions developed in previous work that are linked to the implementation and field-scale simulation study carried out in this work.  }  

{As discussed in Section \ref{intro}, laboratory observations indicate that \co exposure can induce chemical alteration of the native rock. Assuming saline aquifers are initially water-wet, we consider that long-term \co exposure on the order of months can shift wettability to intermediate wet or even \co wet, where each wetting state is defined by the contact angle (CA) between water and solid as shown in the left-most illustrations in Figure \ref{fig:data}. As a result, the saturation functions are altered continuously in time, which is shown schematically for both the phase relative permeability and capillary pressure functions (center Figure \ref{fig:data}). Alteration continues until the wetting condition has reached a final state under maximum exposure.}

{When studying this phenomena at the field scale, we expect that dynamics in saturation functions will be inherently linked to local \co exposure, which itself is controlled by migration of \co at the field scale. As a result, different points of the aquifer will have a different local history of \co exposure, which leads to local alteration of saturation functions that varies in space and time throughout the storage site. This spatial/temporal complexity in capillary pressure and relative permeability can in turn impact field-scale \co migration in ways that are not easily predicted \textit{a priori}. Field-scale studies are needed to quantify the impact of this interplay between dynamics in saturation functions and large-scale flow processes, illustrated in the right-most panels in Figure \ref{fig:data}. }

{The link between \co exposure and time-dependent saturation functions requires a model to evolve capillary pressure and relative permeability functions as the wetting condition continuously changes. As depicted under ``Modeling'' in Figure \ref{fig:data}, we have previously developed new formulations for capillary pressure and relative permeability that incorporate exposure to \co as an additional dynamic variable \citep{Kassa2019, Kassa2020}. These dynamic models are implemented in a standard field-scale simulation approach (described in Section \ref{TPTCmodel}) that includes a brief overview of the dynamic model development from previous work in Section \ref{SatuFunc}.}
  
\subsection{Field-scale governing laws}\label{TPTCmodel}
\co storage in saline aquifers is modeled as a three-dimensional flow system consisting of two immiscible fluids, \co and brine, that are mutually soluble fluids. In this section, we introduce the model equations to describe the TPTC flow in a porous medium. The general form of the TPTC model is presented below, however we introduce some simplifications later during the model implementation (cf. Section~\ref{sec:implementation}). For the sake of brevity, we consider constant temperature $T$ and salinity in this study, but these effects can be included into the model in a straightforward manner.

Let ${\rm \Omega}\in \mathbb{ R}^d$, $d = 1, 2, ~{\rm or }~3$  be a permeable domain  having a Lipschitz continuous boundary ${\rm \partial\Omega}$ and saturated with two-phase fluids that have two components each.  The phases, wetting and non-wetting, are indicated by subscripts $\alpha\in\{w, n\}$ and the two components are represented by index $k \in \{{\rm\co,  water}\}$. The mass conservation law of component $k$ is described by:
\begin{equation}\label{massbalanceCom}
\phi\sum_\alpha\partial_t(\rho_\alpha  S_\alpha X_{\alpha}^k ) +  \sum_\alpha\nabla\cdot( \rho_\alpha X_{\alpha}^k  \vec{u}_\alpha + \vec{j}_\alpha^k) = F^k  ~{\rm in}~ \Omega,
\end{equation}
where $X_\alpha^k$ is component $k$ mass fraction in phase $\alpha$, $\phi$ denotes the porosity of the porous domain, $S_\alpha$ is the phase $\alpha$ saturation, $\rho_\alpha$ is the density of phase $\alpha$, $\vec{u}_\alpha$ is the Darcy flux of phase $\alpha$, and $F^k$ is the source term of \mbox{component $k$.} 

For each phase $\alpha\in \{w,n\}$, the Darcy flux $\vec{u}_\alpha$ is given by the multiphase extension of Darcy's law:
\begin{eqnarray}\label{Darcy}
\vec{u}_\alpha = -\frac{\mathbb{K} k_{r\alpha}}{\mu_\alpha}\big(\nabla P_\alpha - \rho_\alpha \vec{g}\big),
\end{eqnarray}
where  $\mathbb{K} : \mathrm{\Omega} \rightarrow \mathbb{R}^{d\times d}$ is the intrinsic permeability tensor of the rock, $k_{r\alpha}$ is the relative permeability for phase $\alpha$,  $\mu_\alpha$ is phase viscosity, $P_\alpha$ is phase pressure, 
 and $\vec g$ is the gravitational vector. Hereafter, we will use $\mathbb K$ and $K$ for a tensor and scalar permeability, respectively. The diffusive flux of component $k$ in  phase $\alpha$, $\vec{j}_\alpha^k$, is represented by the Fick's law and has the form of:
\begin{equation}\label{Diffusion}
\vec{j}_\alpha^k = -\rho_\alpha \rm D^k_{\alpha}{\tau}_{\alpha} S_\alpha\phi\nabla X_{\alpha}^k,
\end{equation}
where $\rm D^k_{\alpha}$ is the molecular diffusion coefficient and $\tau_{\alpha}$ is tortuosity for phase $\alpha$ which is calculated as \citep{Millington1961}:
\begin{equation}
\tau_{\alpha}=\frac{(\phi S_{\alpha})^\frac{7}{3}}{\phi^2}.
\end{equation}

\noindent The component mass fractions, $X_\alpha^k$,  satisfies:
\begin{equation}\label{mol-fra}
\sum_k X_\alpha^k = 1.
\end{equation}
 
\noindent The phase $\alpha$ saturation takes values between zero and one, and the sum of phase saturation equals one. That is:
\begin{equation}\label{satu}
0\leq S_\alpha\leq 1, ~{\rm and}~\sum_\alpha S_\alpha = 1.
\end{equation}

\noindent The pressures, wetting and non-wetting, are connected by the capillary pressure relation as 
\begin{equation}\label{cap-pressure}
P_c(\cdot, S_w) = P_{n} - P_w.
\end{equation}

\noindent A model for mass transfer between the phases is needed to close the system of equations. In the general form, we consider the fugacity constraints
\begin{equation}\label{fugacity}
f_w^k(P_w,T,X_w^k)-f_{n}^k(P_n,T,X_n^k) = 0, 
\end{equation}
where they express the requirement that non-wetting (gas) and wetting (liquid) fugacities have to be equal for each component \citep{Coats1980,Voskov2012}. The fugacity coefficients are calculated according to the work of \cite{Spycher2005}. We note that other forms of mass transfer may be considered, including equilibrium partitioning. If mass transfer is not modeled, then this closure relation may be omitted from the model formulation.

Equations \eqref{massbalanceCom}  through \eqref{fugacity} with appropriate initial and boundary conditions can be used to describe TPTC flow dynamics in a porous medium. These equations provide a complete description of the physics of isothermal TPTC flow in a porous medium bearing in mind that the relative permeabilities and capillary pressure parameterizations are given.
 
\subsubsection{Previous work: Dynamic saturation functions}\label{SatuFunc}
Capillary pressure in Equation \eqref{cap-pressure} and phase relative permeability in Equation \eqref{Darcy} can be modeled by functions such as \cite{RefBrooksCorry} or \cite{RefvanGenuchten} constitutive functions. However, these standard flow functions are limited to static wetting conditions and lack the time component needed to capture dynamics in constitutive relations due to long-term WA. In recent work, we have developed new dynamic models for both capillary pressure \citep{Kassa2019} and relative permeability \citep{Kassa2020} that capture the underlying pore-scale WA mechanisms caused by exposure to \co. 
 
Below we briefly review the development of the upscaled dynamic flow functions from the pore scale to the Darcy scale performed in previous work. Although the development starts from pore-scale parameters, the final result are flow functions that solely rely on macroscale variables, where the additional dynamic parameters have been calibrated to the pore-scale simulations performed previously. The reader is referred to the referenced papers for more details.
 
The dynamic constitutive models in \cite{Kassa2019} and \cite{Kassa2020} were developed by applying an upscaling workflow that involves correlating Darcy-scale models to dynamic capillary pressure and relative permeability data generated from pore-scale numerical experiments. The key features to this workflow are: 
\begin{enumerate}
     \item A mechanistic model for CA change from an initial to final wetting state that is incorporated directly in pore-scale simulations.
     \item A dynamic model formulation identified at the Darcy scale that incorporates exposure time and can be correlated to the dynamic data with as few parameters as possible.
     \item A link between the correlated dynamic Darcy-scale parameter(s) and the pore-scale model for CA change.
\end{enumerate}

{We note that the above approach does not capture changes in residual saturation for \co and brine as a result of different wetting conditions. Although the connection between residual saturation and wetting condition is well studied, incorporation into a dynamic framework is still a subject of ongoing work. As such, the dynamic models presented are limited, but still provide important insight into \co migration patterns due to changing wetting conditions. }

The first step in the previously published upscaling workflow is the pore-scale mechanistic model for CA change. The chosen model was motivated by experimental evidence that indicates a gradual and permanent change in the fluid-fluid CA when exposed to a WA agent, in this case \co, over a long period of time. CA change is caused by adsorption of the WA agent to the pore surface that subsequently alters surface chemistry and changes the affinity of fluids to the solid. In addition, a longer exposure time induces a greater change in CA. We thus designed a dynamic CA model that is sorption-based and time-dependent and has the form (see \cite{Kassa2019} for development details):
\begin{equation}\label{CA_model}
    \theta = \theta^{\rm i} + \frac{(\theta^{\rm f}-\theta^{\rm i})\chi}{C + \chi},
\end{equation}
where $\theta^{\rm i}$ and $\theta^{\rm f}$ are the initial and final CA respectively, $C$ is a model parameter that determines the rate of WA over time, with increasing $C$ indicating slower CA change, and $\chi$ is a cumulative measure of exposure time to the WA agent at the pore level. The quantity $\chi$ was evaluated by integrating the chosen measure of local exposure, i.e., agent concentration or saturation, over time. As $\chi$ is a quantity that always increases under exposure or remains constant under the absence of the WA agent, the above model leads to irreversible CA change. 

With the above mechanistic WA model incorporated at the pore scale, a set of numerical experiments were performed to simulate laboratory measurements of capillary pressure and relative permeability, i.e., subsequent drainage and imbibition cycles with stepwise changes in saturation over time. The simulated data were obtained by modeling the pore scale as a bundle of cylindrical or triangular tubes. A schematic illustration of capillary pressure and relative permeability data obtained in our previous studies are shown in Figure \ref{fig:data}.

Analysis of the simulated data showed that both saturation functions evolve smoothly from an initial to final state as the CA changes dynamically and heterogeneously with increasing exposure time to the non-wetting fluid. In the reported experiments, CA changed from a strongly water-wet system ($\theta^{\rm i} = 0 ^\circ$) to an intermediate-wet system ($\theta^{\rm f} = 85 ^\circ$), although the opposite WA process could equally have been simulated. Figure \ref{fig:data} shows that capillary gradually decreases over a period of months to years over several subsequent drainage and imbibition processes. Similarly, relative permeability alteration causes an overall decrease in non-wetting phase permeability with increasing exposure time compared with an increase in wetting phase permeability for the same exposure. 

This temporal component motivated a new definition for upscaled exposure time at the Darcy-scale, given as: 
\begin{equation}\label{eqhistflux2}
\overline{\chi}  := \frac{1}{t_{ch}}\int_0^{t}  X^{\rm WA} d\tau,
\end{equation}
where  $t_{ch}$ is a pre-specified characteristic time which is used to scale the history of exposure, and $X^{\rm WA}$ is the chosen measure of exposure of WA agent. Exposure may be either modeled as dissolved mass fraction $X^{\rm WA}=X_w^{\rm \co}$ or as fluid saturation $X^{\rm WA}=S_{n}$, depending on whether the WA agent is a dissolved solvent or invading fluid, respectively. We note that $\overline{\chi}$ is a dimensionless non-linear parameter that depends solely on \co concentration or saturation and time, which are all readily available model variables in Darcy-scale simulators. 

The above definition of exposure time was then used to formulate the dynamic models for capillarity and relative permeability.  In \cite{Kassa2019}, we found that alteration of capillary pressure could be modeled as an interpolation between two static states, where the interpolation coefficient is coupled to macroscale exposure time, $\overline{\chi}$. The resulting formulation is an interpolation-based dynamic capillary pressure model as follows:
\begin{equation}
P_c = \frac{\overline{\chi}S_{ew}}{\beta  + \overline{\chi}S_{ew}} \Big(P_c^{\rm f} - P_c^{\rm i} \Big)+ P_c^{\rm i},  \label{eq:dynPc_nonuni}
\end{equation} 
where $P_c^{\rm i}$ and $P_c^{\rm f}$ are the capillary pressure functions at the initial and final wetting states, respectively, $\overline \chi$ is the exposure time at the Darcy scale, $\beta$ is  a fitting parameter, and $S_{ew}$ is effective saturation which can be defined as:
\begin{equation}
S_{ew} = \frac{S_w - S_{rw}}{1- S_{rw} - S_{rn}},
\end{equation}
where $S_{rw}$ and $S_{rn}$ are wetting and non-wetting residual saturations, respectively. The end wetting-state capillary pressure functions $P_c^{\rm i}$ and $P_c^{\rm f}$ are modeled as static functions that can be represented by a Brooks-Corey model:
\begin{equation}
P_c^{\rm i} = c^{\rm i}S_{ew}^{-1/\lambda} ~ {\rm and} ~ P_c^{\rm f} = c^{\rm f}S_{ew}^{-1/\lambda},
\end{equation}
where $1/\lambda$ is a parameter that represents pore-size distribution, $c^{\rm i}$ and $c^{\rm f}$ are entry pressures for initial and final wetting-state conditions, respectively.

Similar analysis in \cite{Kassa2020} showed that relative permeability dynamics could be more efficiently modeled by including the exposure time directly into the function parameters rather than following the interpolation approach used above. Furthermore, we found a reduced LET relative permeability model is preferable due to its flexibility. More specifically, from the original relative permeability models in \cite{Lomeland2005} (six fitting parameters) we have obtained reduced models (two fitting parameters) based on assumptions in the pore geometry, and we have added a dynamic function to capture the wettability dynamics in the relative permeability-saturation curves. This approach results in a dynamic model for relative permeability as follows:
\begin{equation}\label{relativas}
k_{rw} =  \frac{\mathbb{F}(\overline{\chi})S_{ew}^\Lambda}{1-S_{ew} +\mathbb{F}(\overline{\chi}) S_{ew}^\Lambda},\quad k_{rn} = \frac{1-S_{ew}}{1-S_{ew} +\mathbb{F}(\overline{\chi})S_{ew}^\Lambda},
\end{equation} 
where the dynamic function $\mathbb{F}(\overline{\chi})$ is given by 

\begin{equation}\label{LETdyna}
\mathbb{F}(\overline{\chi}) = \begin{cases}
\eta\overline{\chi} + {\rm E}^{\rm i} &{\rm if}\quad  \overline{\chi}<\frac{E^{\rm f}-E^{\rm i}}{\eta},\\
E^{\rm f} &{\rm if}\quad  \frac{E^{\rm f}-E^{\rm i}}{\eta}\leq \overline{\chi}.
\end{cases}
\end{equation} 

\noindent ${\rm E}^{\rm i}$ and  ${\rm E}^{\rm f}$ are empirical data fitting parameters for the initial and final wetting-state condition and  $\eta$ is a dynamic fitting parameter that controls the WA induced dynamics in the relative permeabilities. 

{The reader is referred to the cited papers for more details on the comparison between the simulated data and the correlated functions. In summary, the correlated models perform well for any general saturation history resulting in an excellent match with the simulated data, thus demonstrating the robustness of the dynamic models. }

We emphasize that the above presentation is intended to provide background on the approach used to understand and model the manifestation of pore-scale WA at the field scale. In this study, only the resulting macroscale dynamic flow functions, Equations \eqref{eq:dynPc_nonuni} and \eqref{LETdyna}, and associated parameters are applied further in field-scale simulations. CA change is not explicitly modeled in these simulations, i.e., $\theta$ is not a variable at the field scale, but the mechanisms of CA change described in Equation \ref{CA_model} are implicitly captured in the dynamic flow functions. This implicit connection between pore scale and macroscale is described further in the next section.

\subsubsection{Dynamic model implementation and parameterization}
The above dynamic models in Equations \eqref{eq:dynPc_nonuni} and \eqref{LETdyna} with associated parameters are substituted directly into the TPTC model in place of the usual static functions, i.e., capillary pressure relation in Equation \eqref{cap-pressure} and the phase relative permeability functions as part of the Darcy flow equation in \eqref{Darcy}. The functions are easily implemented in a standard flow simulator, involving only an additional functional dependency on exposure time, itself a straightforward calculation from Equation \eqref{eqhistflux2} from model variables. 

There are additional parameters in both Equations \eqref{eq:dynPc_nonuni} and \eqref{LETdyna} which need to be obtained by characterization of the rock-fluid system of interest. For instance, capillary pressures at end wetting states needed in \eqref{eq:dynPc_nonuni} can be obtained from laboratory analysis, possibly by rescaling curves obtained by nitrogen or mercury drainage experiments. For relative permeability, the initial wetting state needs to be fully characterized using a non-reactive fluid.

The dynamic $\beta$ and $\eta$ parameters are obtained by characterization of WA, either by fitting to laboratory or numerical pore-scale experiments performed for this purpose. In our analyses in \cite{Kassa2019} and \cite{Kassa2020}, we found a direct connection between the dynamic parameters and the underlying pore-scale WA model for CA change. In both cases, $\beta$ and $\eta$ have a correlated relationship with the pore-scale parameter $C$ in Equation \eqref{CA_model}. 
According to \cite{Kassa2019}, the parameter $\beta$ is represented as a power function of the pore-scale parameter $C$:
\begin{equation}\label{beta-C}
    \beta = b_1C^{b_2},
\end{equation}
where $b_1>0$ and $b_2>0$ are fitting parameters and are determined from WA experiments. As $C$ decreases, indicating faster WA, $\beta$ decreases. If $C=0$,  wettability changes from the initial to final wetting state instantaneously. 

On the other hand in \cite{Kassa2020}, the parameter $\eta$ is related linearly, but inversely, with parameter $C$ as:
\begin{eqnarray}\label{eta-C}
\eta = -\nu_1C + \nu_2,
\end{eqnarray}
where $\nu_1>0$ and $\nu_2>0$ are empirical fitting parameters. Faster WA (decreasing $C$) results in increasing $\eta$, reaching a maximum value, $\nu_2$, when $C=0$.

Equations \eqref{beta-C} and \eqref{eta-C} imply that the dynamic capillary pressure and relative permeability behaviors can be estimated by knowing the CA change at the pore level over exposure time to the WA agent. This could be characterized by relatively straightforward batch exposure experiments under different exposure times, and thus avoiding extensive drainage-imbibition experiments.
 
\subsection{Implementation for macroscale simulation}
\label{sec:implementation}

There are numbers of powerful numerical porous media simulators for TPTC flow scenarios. For instance, DuMux \citep{Flemisch2011}, MRST \citep{Knut2016}, and OPM \citep{Rasmussen2019} are among the software that are used to solve multi-phase flow models in a porous medium. These simulators solve the flow problem by reducing the continuous system of PDEs to an algebraic system of equations. In this paper, we employed the OPM framework to solve and analyze the TPTC model above. 

OPM Flow is a reservoir simulator available in OPM capable of performing TPTC field-scale simulations. This simulator implements a reduced version of the presented TPTC model, where the components in the phases are computed by equilibrium partitioning and molecular diffusion is not included. This simplifications result in faster simulations for advection-dominated problems (e.g., continuous well injection of \co). In studies where it is necessary to include molecular diffusion (e.g., \co migration into caprock), then the complete TPTC model implemented in opm-models can be used, which model the mass transfer between phases by fugacity constraints. We refer to the latter simulator as TPTC simulator.

Originally, both OPM Flow and TPTC simulators were developed for flow systems with standard saturation functions under the assumption of static and uniform  wetting condition in space and time. We have implemented the dynamic saturation functions in both simulators. This is done by adding the dynamic saturation functions (Equations (\ref{eq:dynPc_nonuni}) and (\ref{LETdyna})) and a new variable $\bar{\chi}$ (Equation \ref{eqhistflux2}) to the simulator, whose value is approximated by the weighted (time step $\Delta t$ over characteristic time t$_{ch}$) cumulative sum of the measure of exposure of WA agent ($S_n$ or $X_{w}^{\rm CO_2}$). 

The OPM version which is used in this study is based on the 2021-04 Release. The two point flux approximation (TPFA) and backward Euler (BE) method are used to discretize the model in space and time, respectively. The resulting system of equations is linearized using the Newton-Raphson method. The simulator uses automatic-differentiation (AD) to calculate the Jacobian of the system. We refer to \cite{Rasmussen2019} for details on the implementation and model setup of Flow. The links to download the corresponding code to reproduce the numerical studies can be found at the end of the manuscript.

\section{Numerical experiments and results}\label{NumMeth}

In this section, we apply the model and implementation in Section 2 to examine the effect of long-term WA on \co plume migration in the reservoir and containment by the caprock. We consider a change in wettability between two wetting states that is caused by exposure to WA agent, which results in alteration of the saturation functions as described earlier. To understand the important factors affecting long-term WA, we consider different parameters that characterize the fluid-rock system, including intrinsic properties, $\rm K$, $\phi$, $c^{\rm i}$, flow rate $q$, in combination with different values of the WA dynamic coefficients, $\beta(C)$, and $\eta(C)$. 

The impact is investigated by simulating \co migration under WA for two simple systems: (1) one-dimensional horizontal flow system (1D-H) of constant \co injection and (2) one-dimensional vertical flow system (1D-V) of \co placed under a caprock that is initially sealing. These examples give insight into the impact of WA on storage efficiency and storage containment. Finally, scaling relationships are proposed to understand the parameters for which the impact of WA is significant. 

\subsection{End-wetting state saturation functions}\label{sec:Bl}

\begin{figure}[h!]
    \centering
    \subfigure{\includegraphics[scale=0.42]{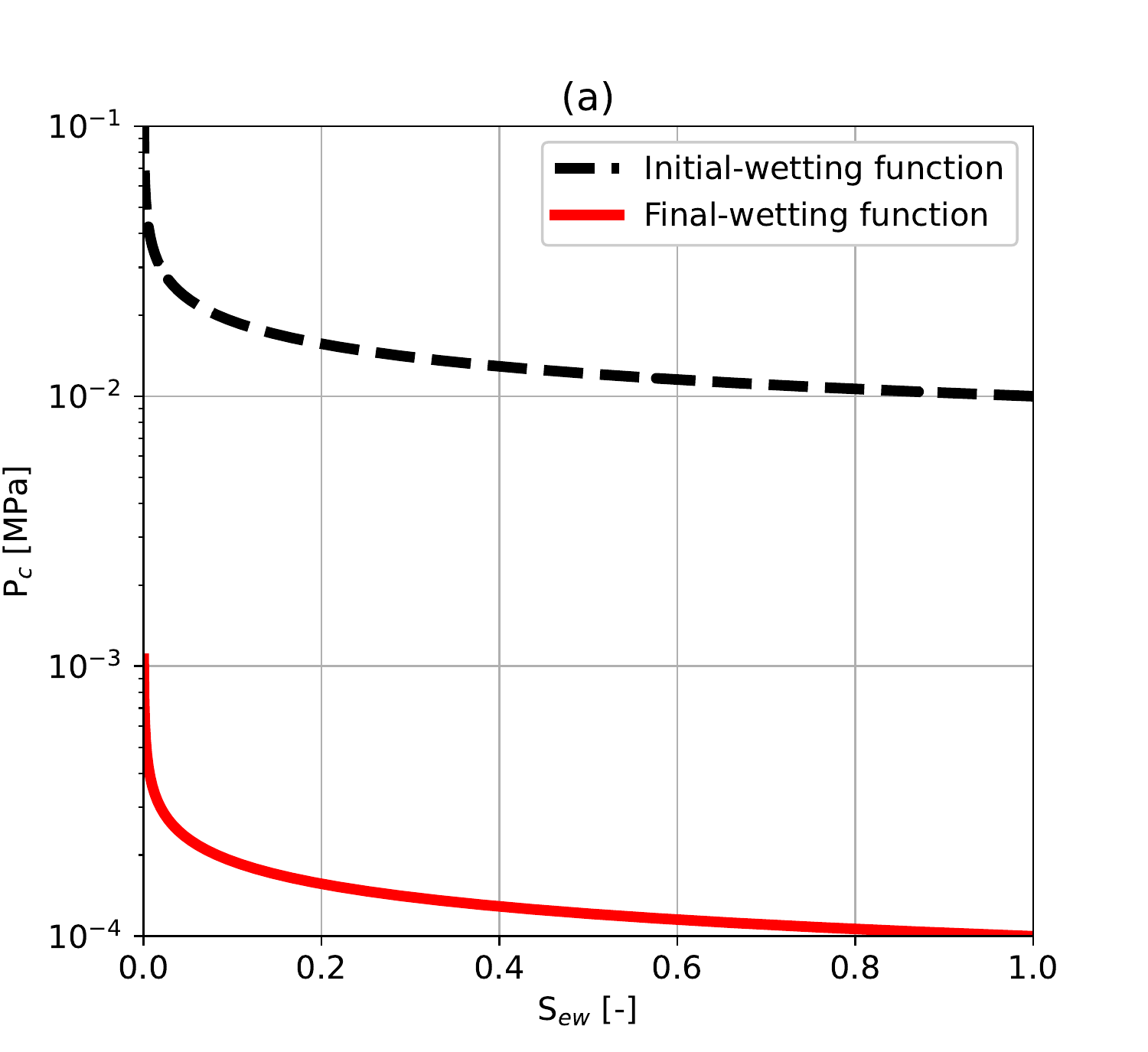}}
    \subfigure{\includegraphics[scale=0.42]{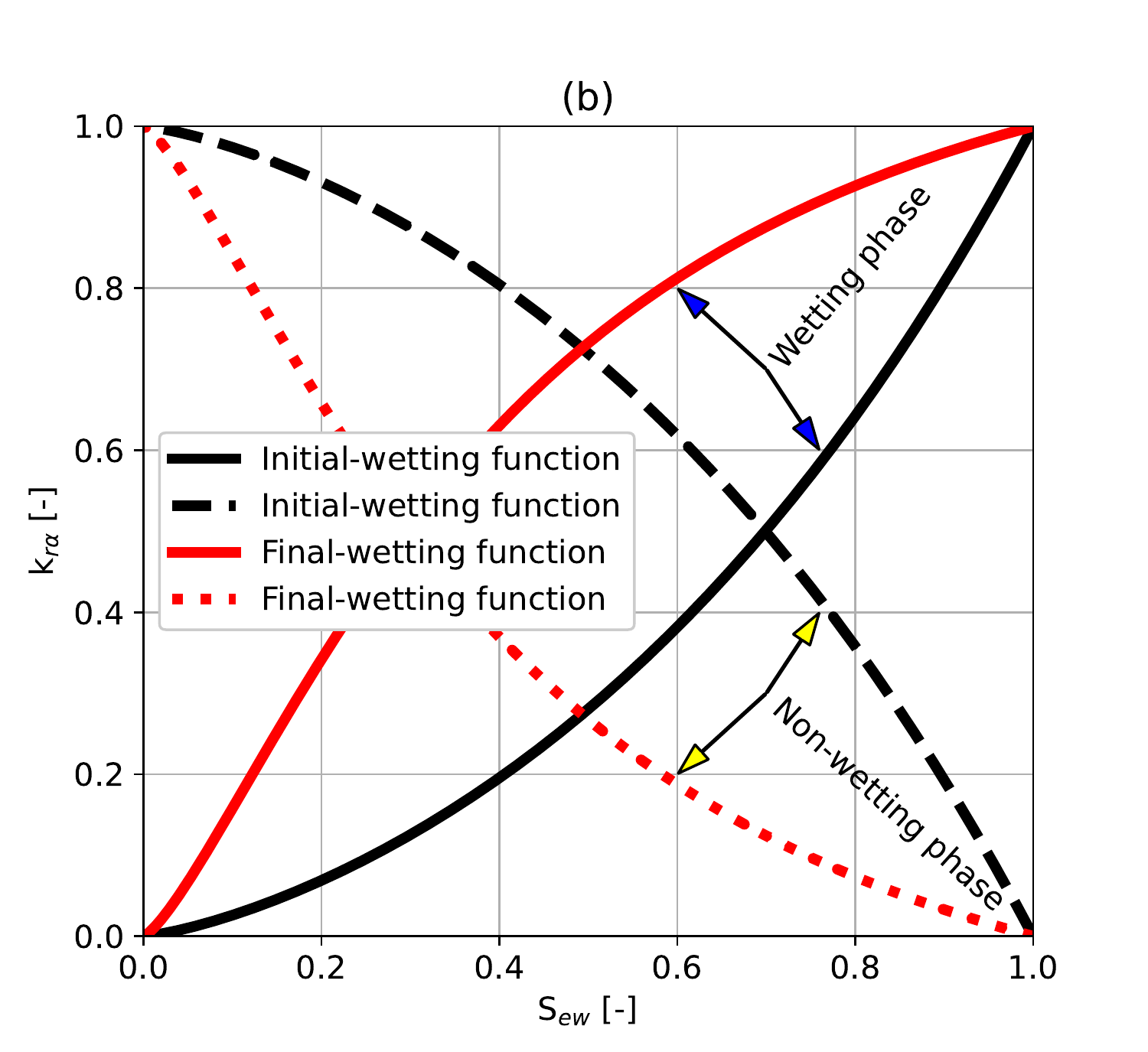}}
   \subfigure{\includegraphics[scale=0.42]{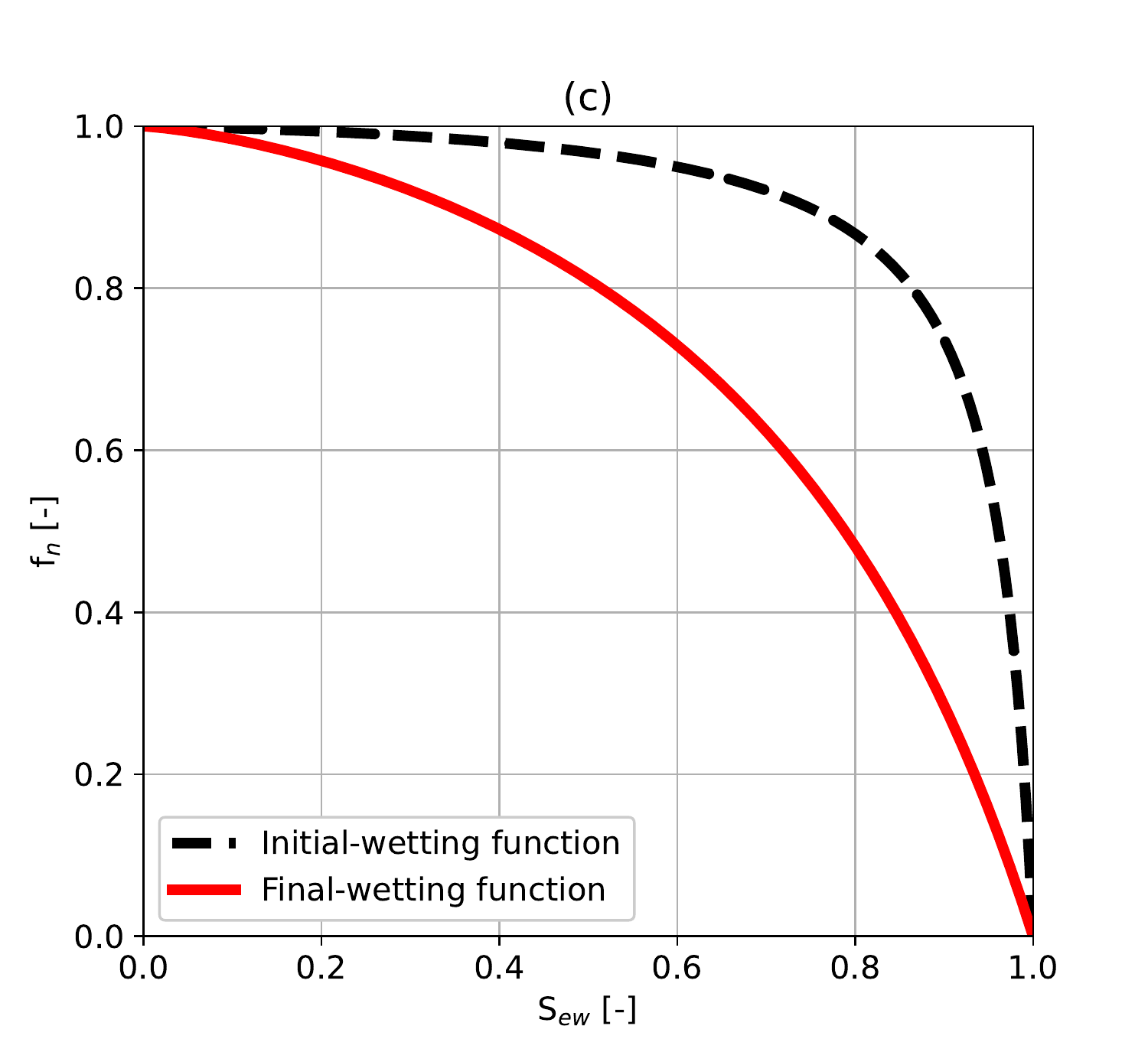}}
    \subfigure{\includegraphics[scale=0.42]{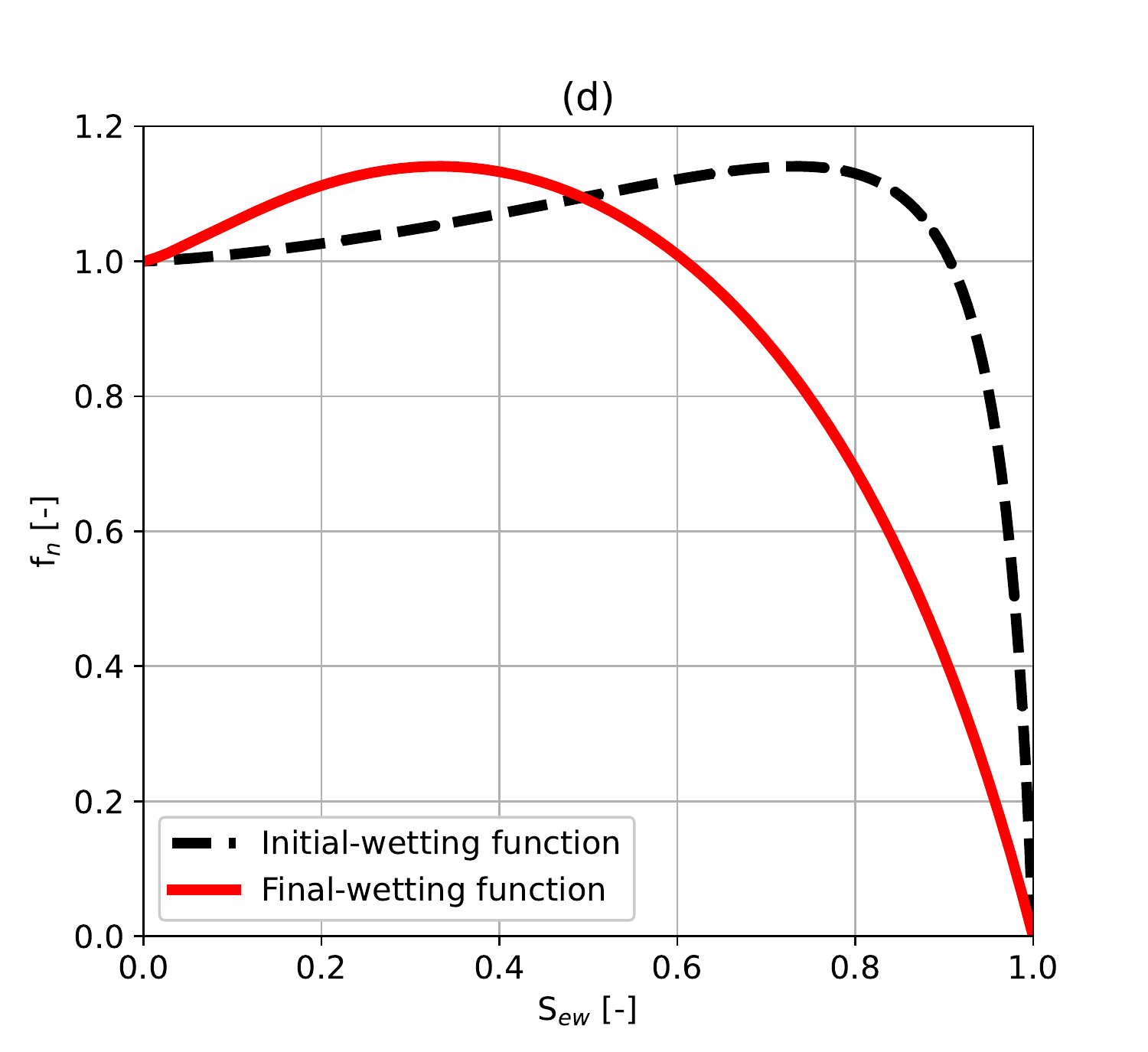}}
    \caption{End-wetting saturation functions: (a) capillary pressure, (b) phase relative permeability, (c) fractional flow function for horizontal flow, and (d) fractional flow functions with gravity effect (using the parameters in Table \ref{BtubeM_End} and $q=10^{-3}$ m$^3$/s).}
    \label{fig:fracflow}
\end{figure}

To start, we present the saturation functions applied in the reservoir for the initial- and final-wetting states. These functions form the basis of the dynamic models for both capillary pressure (Equation~\ref{eq:dynPc_nonuni}) and phase relative permeability (Equation~\ref{relativas}) presented earlier. 
The model parameters used in this study are given in Table \ref{BtubeM_End}, and the associated curves are depicted in Figures \ref{fig:fracflow}a and \ref{fig:fracflow}b.  We note that the chosen parameter values for the saturation functions are not measured values but are consistent with values obtained under static wettability conditions in our previous work \citep{Kassa2019, Kassa2020}. {However, any end-state saturation functions can be used if appropriate field or lab data are available. }
\setcounter{table}{0}
\begin{table}[h!]
\caption{Table of model parameters for the numerical studies.}\label{BtubeM_End}
\centering
\begin{tabular}{ l l l l}
\hline
Symbol   & Value  & Unit      \\
\hline 
$c^{\rm i}$        & 10$^{4}$                 & Pa\\ 
$c^{\rm f}$      & 10$^{2}$                & Pa \\
$\lambda$    & 3.6                  & $[-]$   \\  
$\Lambda$    & 1.3                   & $[-]$ \\   
$E^{\rm i}$   & 0.48                & $[-]$   \\ 
$E^{\rm f}$    & 3.37                    & $[-]$   \\
$\rho_{n}$   & 716.7                 & kg/m$^3$\\
$\mu_{n}$    & $5.916\times 10^{-5}$ & Pa$\cdot$s     \\
$\rho_w$     & 1050                  & kg/m$^3$  \\ 
$\mu_w$      & $6.922\times 10^{-4}$ & Pa$\cdot$s       \\
$K$ & 10$^{-10}$                 & m$^2$\\ 
$S_{rw}$ & 0.2 & $[-]$ \\ 
\hline
\end{tabular}
\end{table}

\noindent We observe in Figure \ref{fig:fracflow}a that WA reduces the entry pressure to the non-wetting phase by two orders of magnitude, causing the capillary pressure function to shift downward while maintaining the same curvature. For relative permeability (Figure \ref{fig:fracflow}b), WA dramatically decreases \co mobility while having an opposite impact on the water mobility. The reduction of \co relative permeability is explained by \co preferring smaller pores with reduced entry pressure, which reduces the ease at which \co can flow. On the other hand,  larger pores become the preferred flow path for water, leading to higher water relative permeability with intermediate-wet/hydrophobic conditions.  {We note that the relative permeability curves in Figure \ref{fig:fracflow}b are consistent with our previous investigation using a bundle-of-tubes approach to upscale the dynamics in saturation functions. These curves exhibit a concavity in the \co relative permeability under water-wet conditions that is not usually observed in the lab. However, we choose these curves in order to explore the impact of a large change in relative permeability curves due to WA even if a realistic system may have a much more narrow alteration.  }
 
We further analyze the static functions in order to understand the flow system before adding dynamics. To this end, it is useful to describe the predominant mechanisms affecting horizontal and vertical flow by way of the fractional flow function (neglecting capillary pressure):
\begin{equation}\label{fracFlow}
f_{n} =\frac{1+\frac{{K}}{q}\lambda_w\Delta\rho g\sin \vartheta }{1 + \frac{\lambda_w}{\lambda_n} } ,
\end{equation}
where $\lambda_{\alpha}$ are the phase mobilities defined as $k_{r\alpha}/\mu_{\alpha}$, $\Delta \rho=\rho_w - \rho_{n}$, $q$ the total flow rate, and $\vartheta$ is the flow inclination angle.  Note that the fractional flow in Equation \eqref{fracFlow} describes horizontal and vertical flow when $\vartheta=0^\circ$ and $\vartheta=90^\circ$, respectively.

 The fractional flow functions are calculated with the values in Table \ref{BtubeM_End} and depicted in Figures \ref{fig:fracflow}c and \ref{fig:fracflow}d. We make several observations based on Buckley-Leverett analysis, which is conducted under the assumption of zero capillary pressure for simplicity. When assuming a constant total flow in one dimension, neither end-wetting state will develop a shock as there is no inflection point in the flow functions. {The development of a shock would be expected if the chosen relative permeability functions were convex as often observed for \co-brine systems per the discussion above.} However, we note that the \co-water front will always advance further for the initial-wet conditions compared to the final-wet conditions, which applies for both horizontal and vertical flow. 

As these curves neglect capillarity, their saturation profile under horizontal and vertical flow will be affected by capillary pressure. Capillarity will smooth saturation shocks and retard the displacing fluid front, and this impact will be greater under initial wet conditions given a stronger capillarity. We also note that constant total velocity is also quite restrictive, but nevertheless gives important insight into flow behavior due to changes in relative permeability and provides a good complement to numerical simulations. 

\subsection{1D-horizontal (1D-H) flow system }\label{1dSimExa}

In this example, we consider a 1600-m homogeneous 1D flow system discretized into 640 equal-sized cells. The system is initially saturated with brine, and \co is injected from the left boundary at a constant rate for one year. The saturation functions are dynamically varied according to the end-wetting states with parameters given previously in Table \ref{BtubeM_End}. In this section, we perform all simulations using the dynamic wettability implementation in the OPM Flow simulator. Here, we neglect the effect of molecular diffusion term and consider the WA agent $X^{\mbox{WA}}$ to be the non-wetting saturation $S_n$.

In \cite{Kassa2019,Kassa2020}, we have found mathematical relationships between the pore-scale parameter $C$ and the dynamic parameters $\beta$ and $\eta$ given in Equations \eqref{beta-C} and \eqref{eta-C}. Based on \cite{Kassa2019,Kassa2020}, for the simulations we consider the pore-scale parameter $C$ in the interval $[10^{-5}, 10^{-4}]$. The parameters $b_1$, $b_2$, $\nu_1$, and $\nu_2$ are estimated by comparing to WA experiments. In this work, we set the values of these parameters to study their dynamical impact in a time scale of order of months ($t_{ch}=10^7$ s). For the simulations in the 1D-H study, these values are shown in Table \ref{BtubeM}. Then, the combination of values $\beta$ and $\eta$ are uniquely determined by the value of $C$. We remark that we use the value of the exponent parameter $b_2$ as calibrated in \cite{Kassa2020}.

\begin{table}[h!]
\caption{Parameters describing the relationship between the pore-scale parameter $C$ associated with CA change and dynamic coefficients $\beta$ and $\eta$ for the 1D-H studies ($t_{ch}=10^7$ s, $X^{WA}=S_n$).}\label{BtubeM}
\centering
\begin{tabular}{ c c c c c c c}
\hline
$b_1$   & $b_2$  & $\nu_1$ & $\nu_2$ & C  & $\beta=b_1C^{b_2}$ & $\eta=-\nu_1 C+\nu_2$      \\
$[-]$   & $[-]$  & $[-]$ & $[-]$   & $[-]$  & $[-]$  & $[-]$           \\
\hline 
&&&&$10^{-5}$ & 1 & 45 \\
&&&&2.5$\times 10^{-5}$ & 5.2 & 37.5 \\
$10^{9}$ &   1.8     &    $4.999\times 10^{5}$    &  $50$ &$5\times10^{-5}$& 18.1 & 25\\
&&&&7.5$\times10^{-5}$& 37.6 & 12.5\\
&&&&$10^{-4}$& 63.1 & $10^{-2}$\\
\hline
\end{tabular}
\end{table}

\subsubsection{1D-H base-case scenario study}
 
\begin{figure}[h!]
\centering
\subfigure{\includegraphics[scale = 0.48]{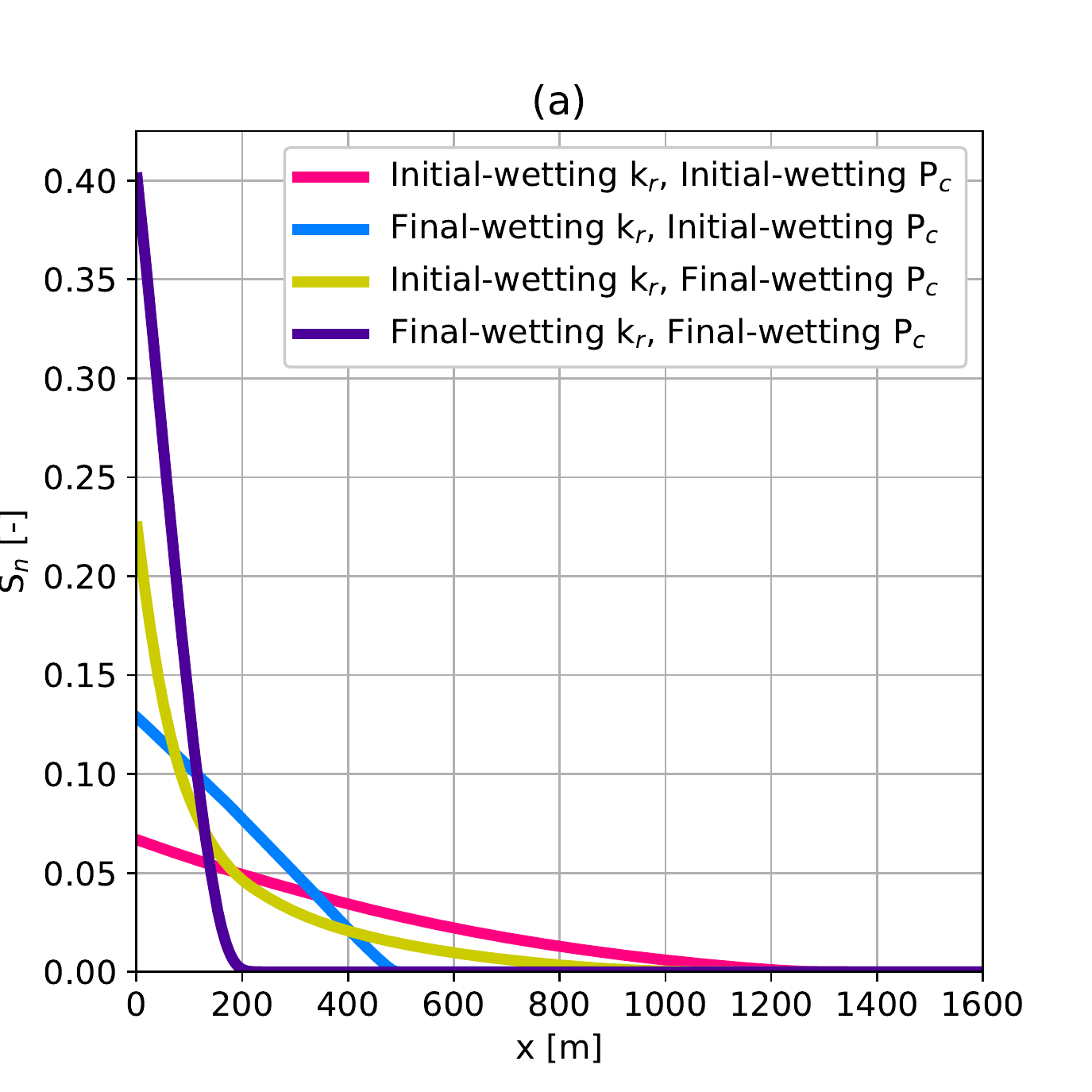}}
\subfigure{\includegraphics[scale = 0.48]{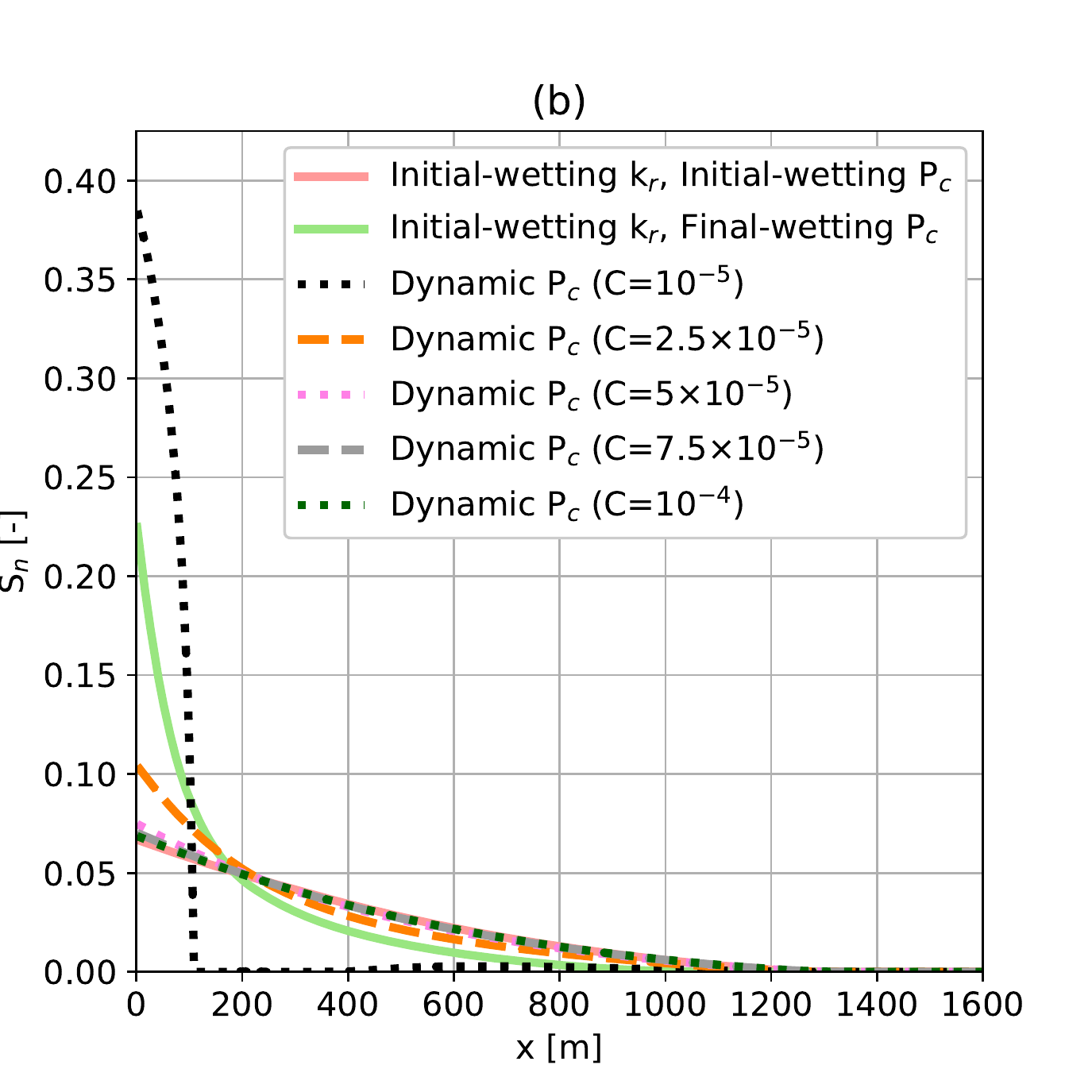}}
\subfigure{\includegraphics[scale = 0.48]{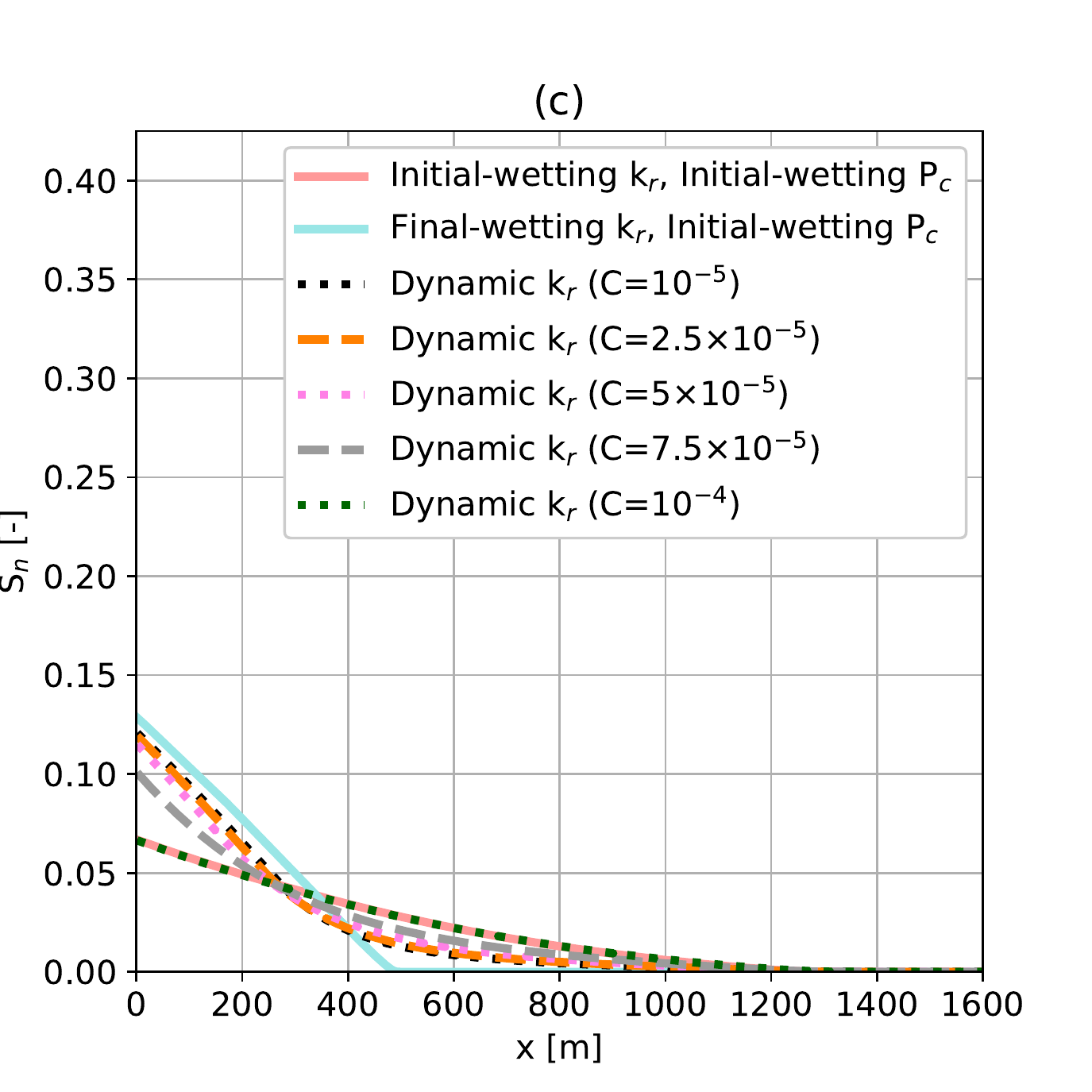}}
\subfigure{\includegraphics[scale = 0.48]{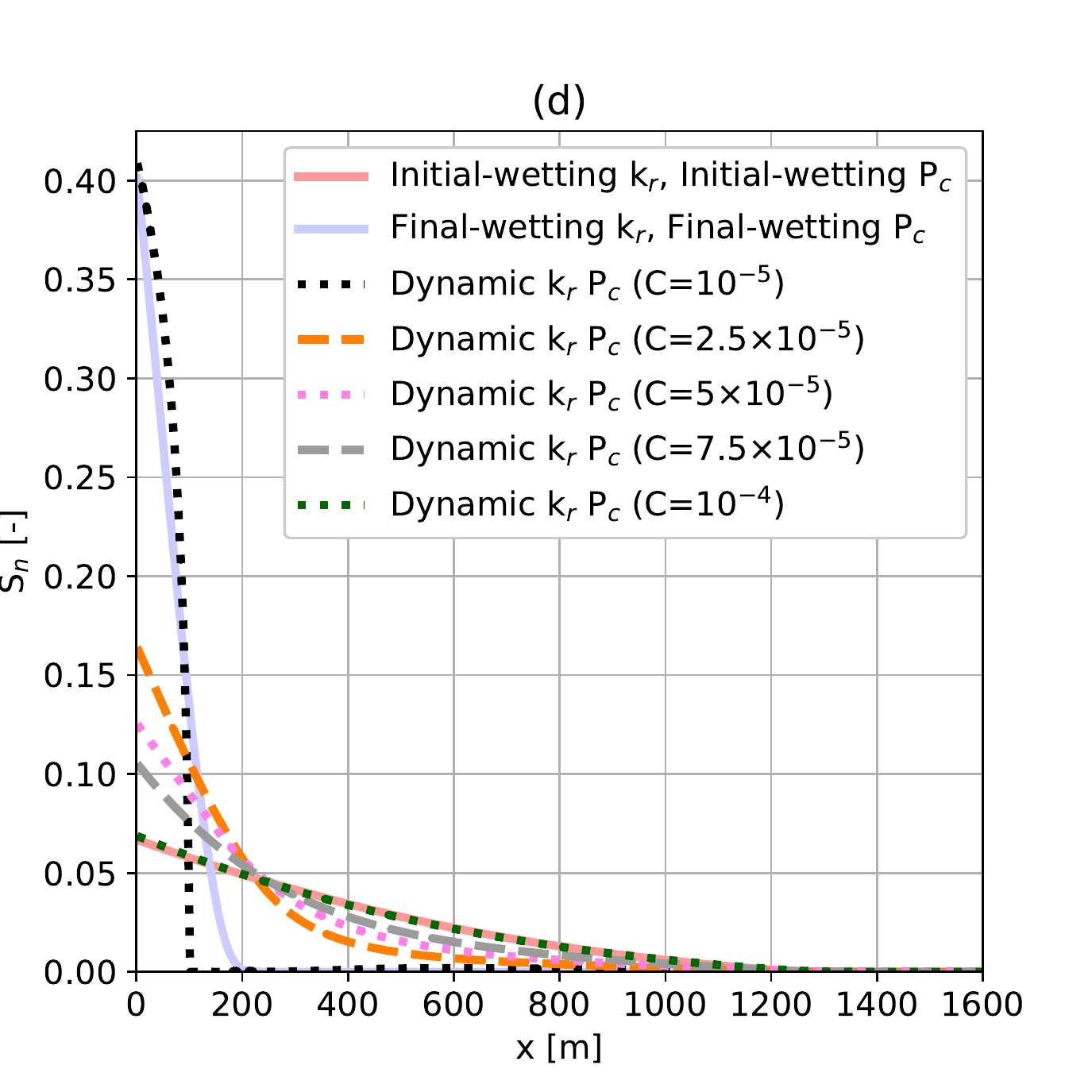}}
\caption{1D-H \co saturation profiles at one year comparing (a) all combinations of initial- and final-wetting functions, (b) different values of the dynamic capillary pressure, (c) different values of the dynamic relative permeability functions, and (d) both dynamic capillary pressure and relative permeability as given in Table \ref{BtubeM}.}\label{horizontal}
\end{figure}

We begin with a base-case scenario where \co is injected for one year for a fixed injection rate, permeability, and porosity, namely $q=10^{-7}$ m$^3$/s, $K=10^{-10}$ m$^2$, and $\phi=0.1$. First, we examine the base-case scenario under static wetting conditions at the two end states. Figure \ref{horizontal}a shows \co migrates far into the domain at low saturation when both functions are in their initial state. Conversely, \co migrates more slowly and fills more of the pore space behind the front when both functions are in their final-wetting states. These results reflect the fractional flow characteristics observed earlier in Figure \ref{fig:fracflow}, but now with the added effect of capillarity. We also observe an independent influence of the wetting state of each saturation function. For instance, we observe the impact of different wetting-state relative permeability functions is more significant when the $P_c$ function is weaker (final-wetting) than for the initial-wetting $P_c$ function.

Next, we investigate the isolated impact of wetting dynamics in capillary pressure by varying only the dynamic WA parameter $\beta(C)$ according to the values in Table~\ref{BtubeM} for the base case scenario. Here, the relative permeability model is kept static at the initial-wetting state. Figure \ref{horizontal}b shows that a small $C$ is required in order for the dynamics in $P_c$ to have an observable impact during the one-year simulated time period, where we recall that smaller $C$ implies faster WA dynamics in $P_c$. In addition, the power-law relationship $\beta(C)$ results in a non-linear dependence on $C$. An important observation can be made regarding the comparison between fastest WA case $C=10^{-5}$ and the reference static case using the initial-wetting $k_r$ and final-wetting $P_c$. One would expect that the dynamic case should be very similar to the static case since the WA dynamics are fast, but here we see that the \co front migrates more slowly and builds to a higher saturation when dynamics are included. This difference indicates that the development of heterogeneous wettability along the horizontal column created by WA dynamics affects \co flow in a complex way that can not be predicted by static wetting simulations alone. 

For the isolated impact of dynamics in relative permeability (Figure \ref{horizontal}c), we test the same values of $C$ that result in the values of $\eta$ listed in Table~\ref{BtubeM}. In these simulations, the capillary pressure model is kept static at the initial-wetting state. Since the model for $\eta(C)$ is linear with $C$, the resulting change in \co migration is more gradual with $C$ than observed for dynamics in capillarity alone. We also observe that the fast-dynamics case ($C=10^{-5}$) does not match the reference static end-state case (final-wetting $k_r$, initial-wetting $P_c$). The \co front in the dynamic case approaches the reference static case towards the inlet (where the wettability has mostly reached the end state), but the front is significantly more advanced where the wettability is still in a transition between initial and final states. 

Combining the impact in both $\beta$ and $\eta$ simultaneously (Figure~\ref{horizontal}d), the results show that the most $C$ cases with slower wettability dynamics remain close to the reference case with static wettability (initial $k_r$, initial $P_c$), which shows the impact of WA dynamics is relatively low given the base-case parameters. The saturation at the inlet begins to build evenly for increasing dynamics reflecting the altered wetting state there, but it is only for the fastest WA dynamics $C=10^{-5}$ where we observe a significant alteration of the location of the \co front. Here again, there is a discrepancy between the fast dynamic case and the final reference static case (final $k_r$, final $P_c$), showing the complex flow behavior when wettability is varying in space and time along the flow path. 

\begin{figure}[h!]
\centering
\includegraphics[scale = .55]{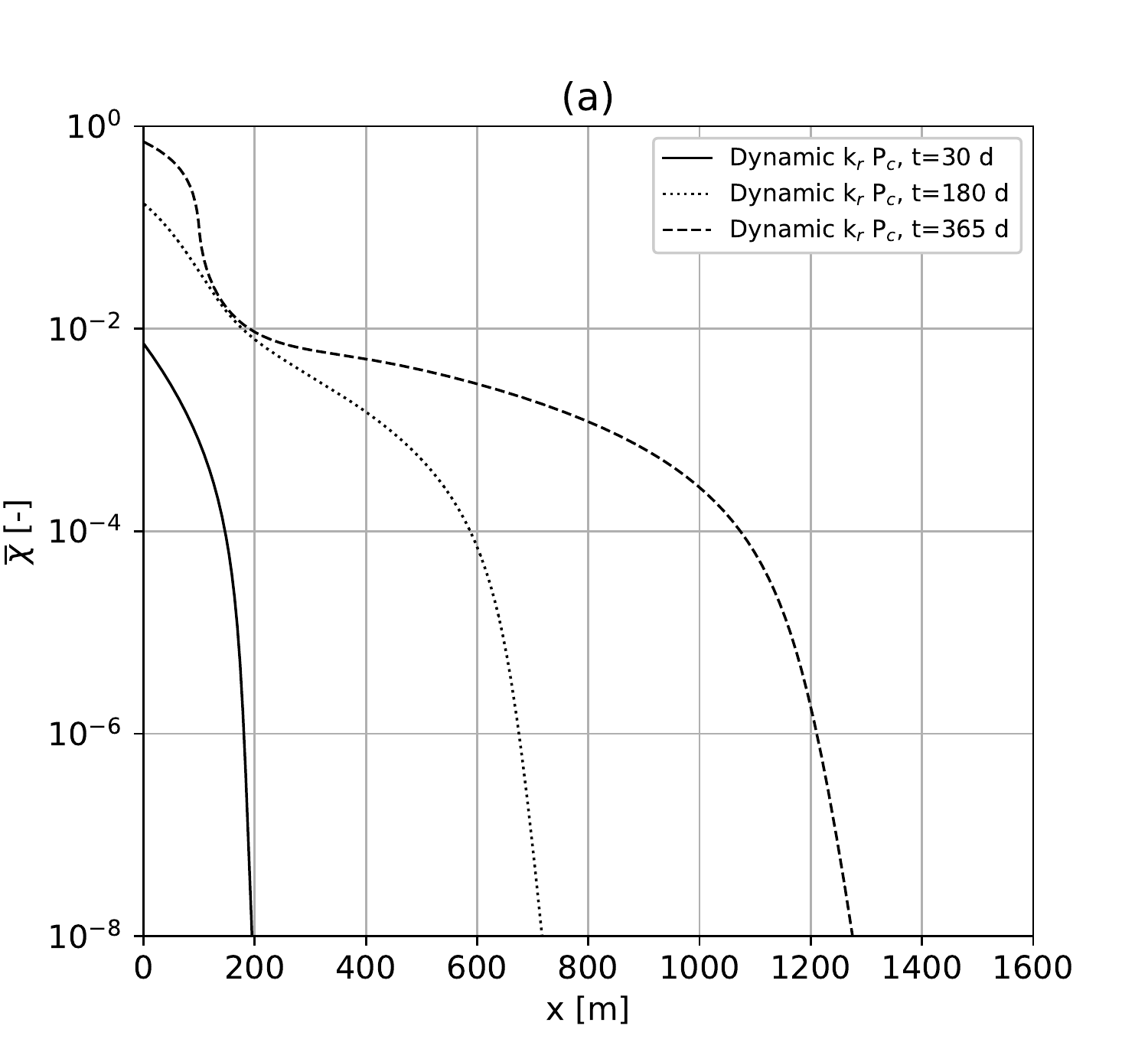}
\includegraphics[scale = .55]{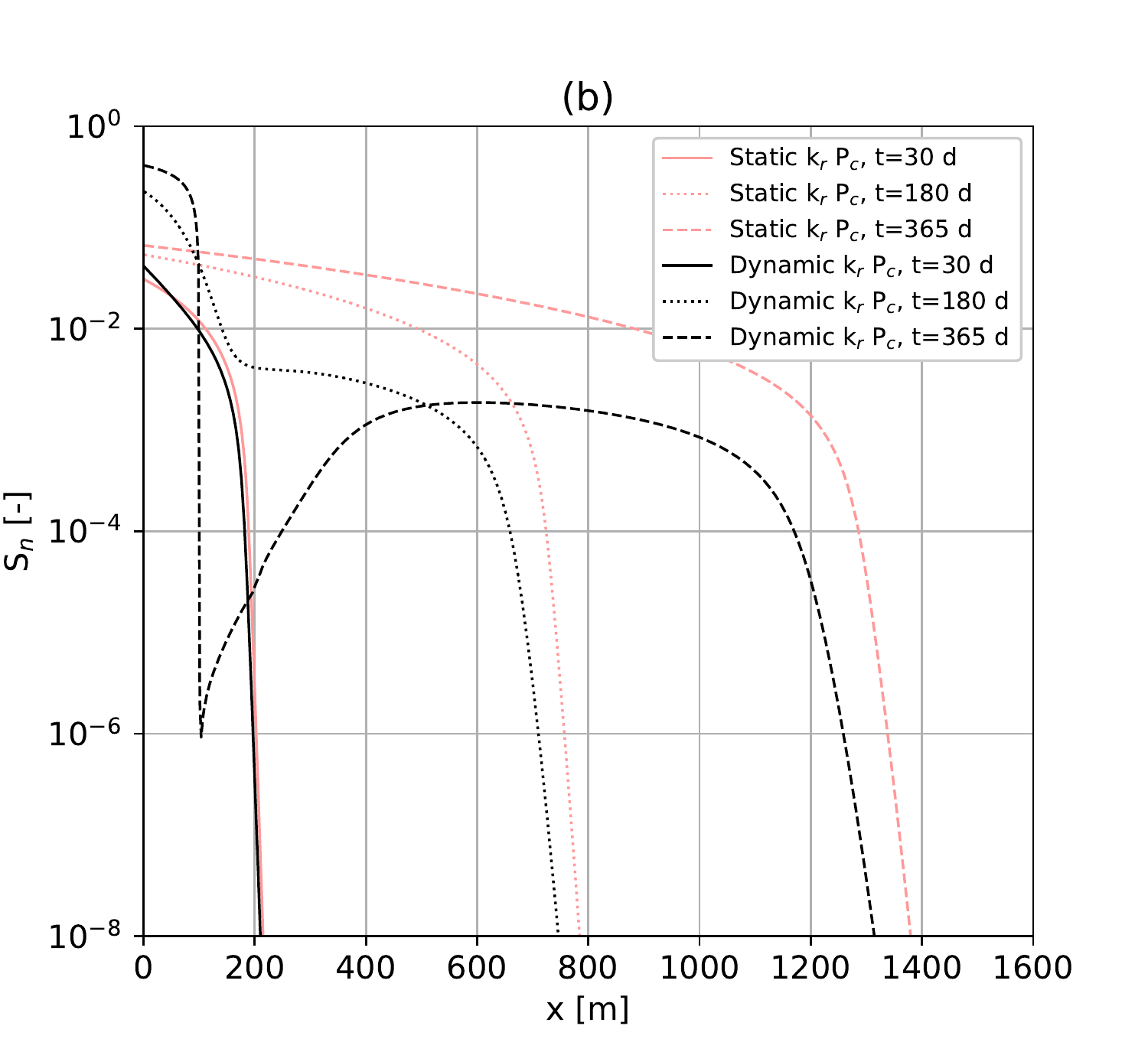} 
\caption{1D-H (a) upscaled exposure time and (b) non-wetting saturation along the aquifer at three different times for the dynamic case $C=10^{-5}$ in Figure~\ref{horizontal}d .}\label{wa_sn}
\end{figure}

Figure~\ref{wa_sn} shows the spatial evolution of the upscaled exposure time $\overline{\chi}$ and the non-wetting saturation $S_n$ for the dynamic case $C=10^{-5}$ in Figure~\ref{horizontal}d at three different injection times. We observe that after 30 days, the evolution of the dynamic $S_n$ profile is practically the same as the static one. This is expected as the wettability alteration effects start to significantly impact the system after a few months, as observed on the saturation profiles at 180 days and one year respectively. These figures confirm the complex flow behavior when fast dynamics are presented.

\subsubsection{1D-H capillary scaling}
We now compare simulations where rock properties ($K$, $\phi$), injection rate ($q$), and the pore-scale parameter ($C$) are considered to vary independently, as given in Table \ref{Param_1d_com}. For the purpose of quantifying the impact of WA dynamics in $P_c$ and $k_{r\alpha}$ compared to a static-wet system, we define the scaled front location difference, SFLD, as:
\begin{eqnarray}\label{sfldcappress}
{\rm SFLD} = \frac{x^{\rm i}-x^{\rm dy}}{x^{\rm i}},
\end{eqnarray}
where $x^{\rm i}$ and $x^{\rm dy}$ denote the \co front location using the static initial-wetting functions and dynamic-wetting functions, respectively. The values of $x^{\rm i}$ and $x^{\rm dy}$ are considered as the further spatial location from the injection well of the non-wetting saturation value greater than a threshold value (here $10^{-4}$). We avoid $x^{\rm i}$ reaching the boundary by considering smaller values of $q$, $K$, and larger values of $\phi$ as observed in Table \ref{Param_1d_com}.

\begin{table}[h!]
\caption{Tested parameter values used in the 1D-H system. A set of simulations were performed for each individual value of $q$, $K$, and $\phi$ in which $C$ is varied, where the combined impact on dynamic wettability functions is studied. For each value of $C$, the corresponding value of $\beta$ and $\eta$ is given in Table \ref{BtubeM}.}\label{Param_1d_com}
  \centering
    \begin{tabular}{c c c c c c}
        \hline
        $q$ & $K$ & $\phi$ & $C$  \\
        m$^3/$s &  m$^2$ &  $[-]$ & $[-]$ \\
        \hline
        $10^{-7}$ & $10^{-10}$   & 0.1   & $10^{-5}$\\
        5$\times 10^{-8}$ &$5\times10^{-11}$   & 0.2   & $2.5\times10^{-5}$ \\
        2.5$\times 10^{-8}$ &$2.5\times10^{-11}$  & 0.4  &$5\times10^{-5}$ \\
        &  &  &$7.5\times10^{-5}$ \\
        &  &  &$10^{-4}$ \\
    \hline
    \end{tabular}
\end{table}
\noindent In addition, we employ a macroscale definition of capillary number, $\mathcal{C}a$, in order to characterize the impact of WA with respect to the flow regime for each parameter combination (Table \ref{Param_1d_com}). Following \cite{Armstrong2014}, we compute $\mathcal{C}a$ on the injection side using the initial entry pressure $c^{\rm i}$ as:
\begin{eqnarray}\label{CapNum}
\mathcal Ca = \frac{\mu_{n}qA\Delta x}{{\phi K} c^{\rm i}},
\end{eqnarray}
where $\Delta x$ is the length of the grid cell at the entry, $A$ the transversal area, and other parameters are defined previously. The transversal area for these simulations is $A$=1 m$^2$. We perform a similar sequence of numerical experiments, first isolating the impact of dynamics in capillarity and relative permeability by allowing dynamics in one function while keeping the opposite function static in the initial-wetting state. Then, we perform experiments of the combined dynamics. The capillary number for these simulations over more than an order of magnitude from $10^{-5}$ to over $10^{-4}$. For reference, the base-case scenario described above has a capillary number of approximately $5\times10^{-5}$.

\begin{figure}[h!]
\centering
\includegraphics[scale = .7]{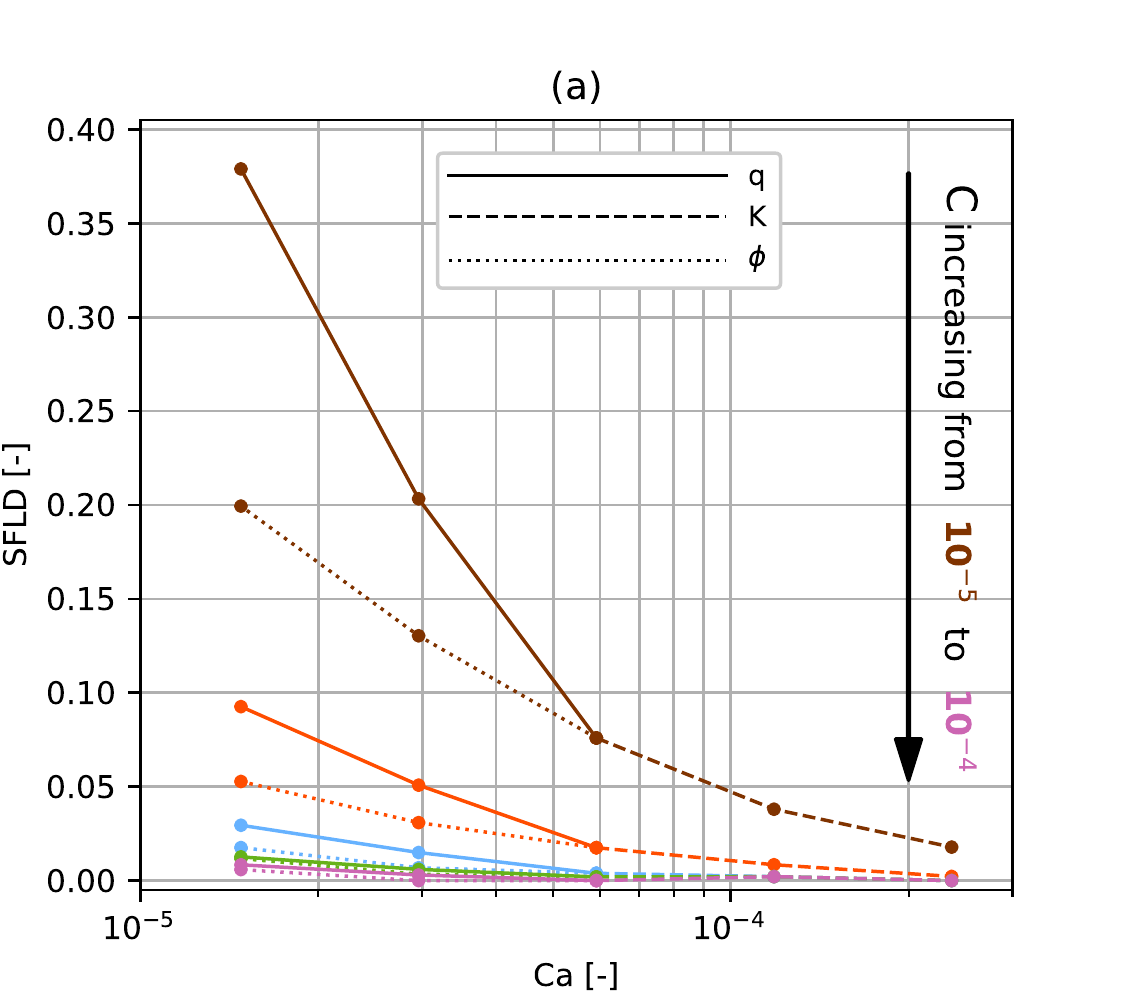}
\includegraphics[scale = .7]{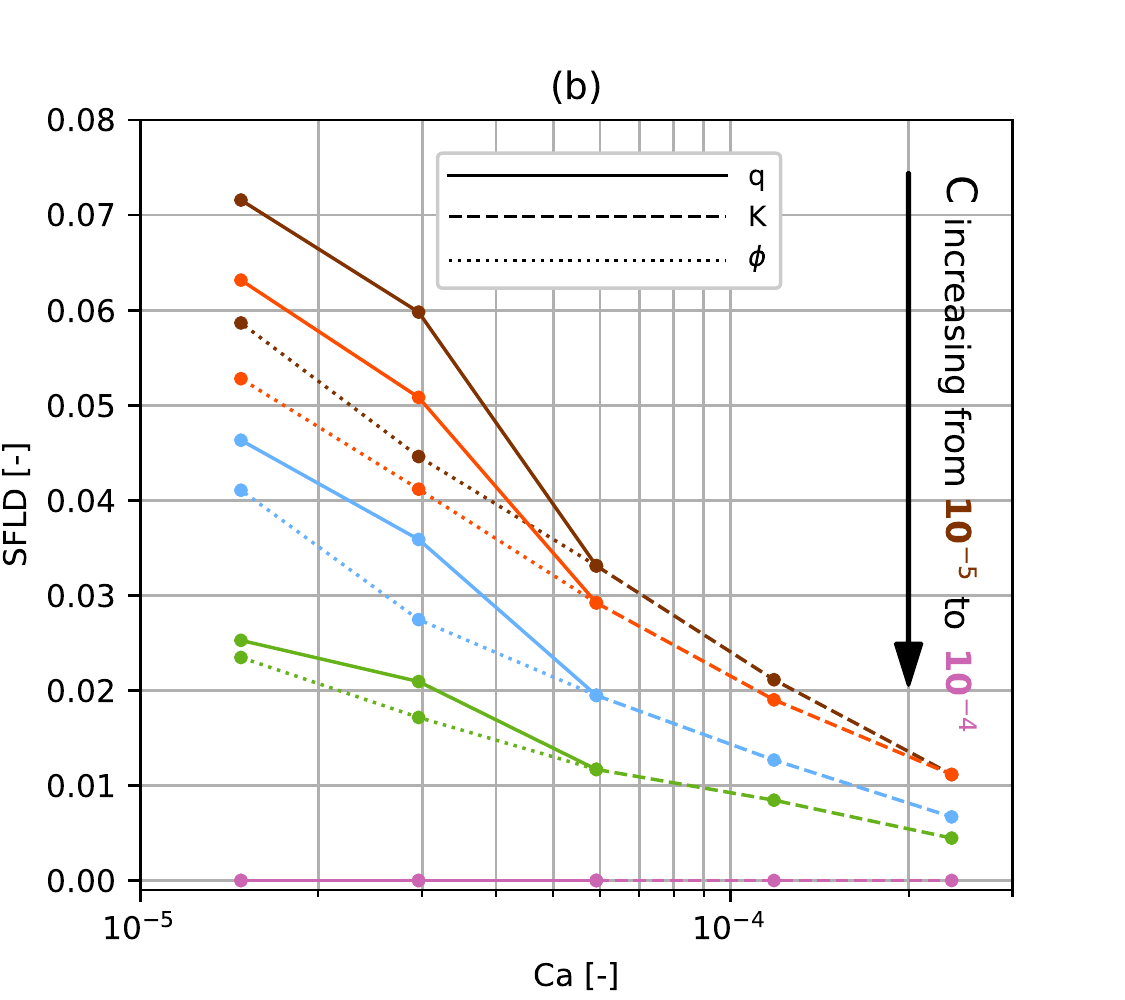} 
\caption{1D-H scaled \co front location difference (SFLD) as a function of capillary number ($\mathcal Ca$) for all simulated parameter combinations (see Table \ref{Param_1d_com}) under isolated dynamics in saturation functions: (a) capillary pressure and (b) relative permeability.}\label{SFLD_Ca}
\end{figure}
 
\noindent We observe the relationship between SFLD and $\mathcal{C}a$ converges onto a single curve for a given $C$ when considering the isolated dynamics in capillary pressure (Figure \ref{SFLD_Ca}a) and relative permeability (Figure \ref{SFLD_Ca}b). This implies that the flow regime influences the impact of WA dynamics. Noting the scale difference on the y-axis for both plots, it is evident that dynamics in $P_c$ have a much greater effect than dynamics in $k_r$ especially at low capillary number, slowing the \co front migration by as much as 35\%.  When $C$ is large (fast WA dynamics) where there is only a 7\% reduction in front location. However, the influence of capillarity diminishes more quickly with higher $\mathcal{C}a$ and higher $C$, and the relative permeability dynamics have a slightly larger impact that is sustained for higher capillary number.  

\begin{figure}[ht!]
\centering
\includegraphics[scale=0.7]{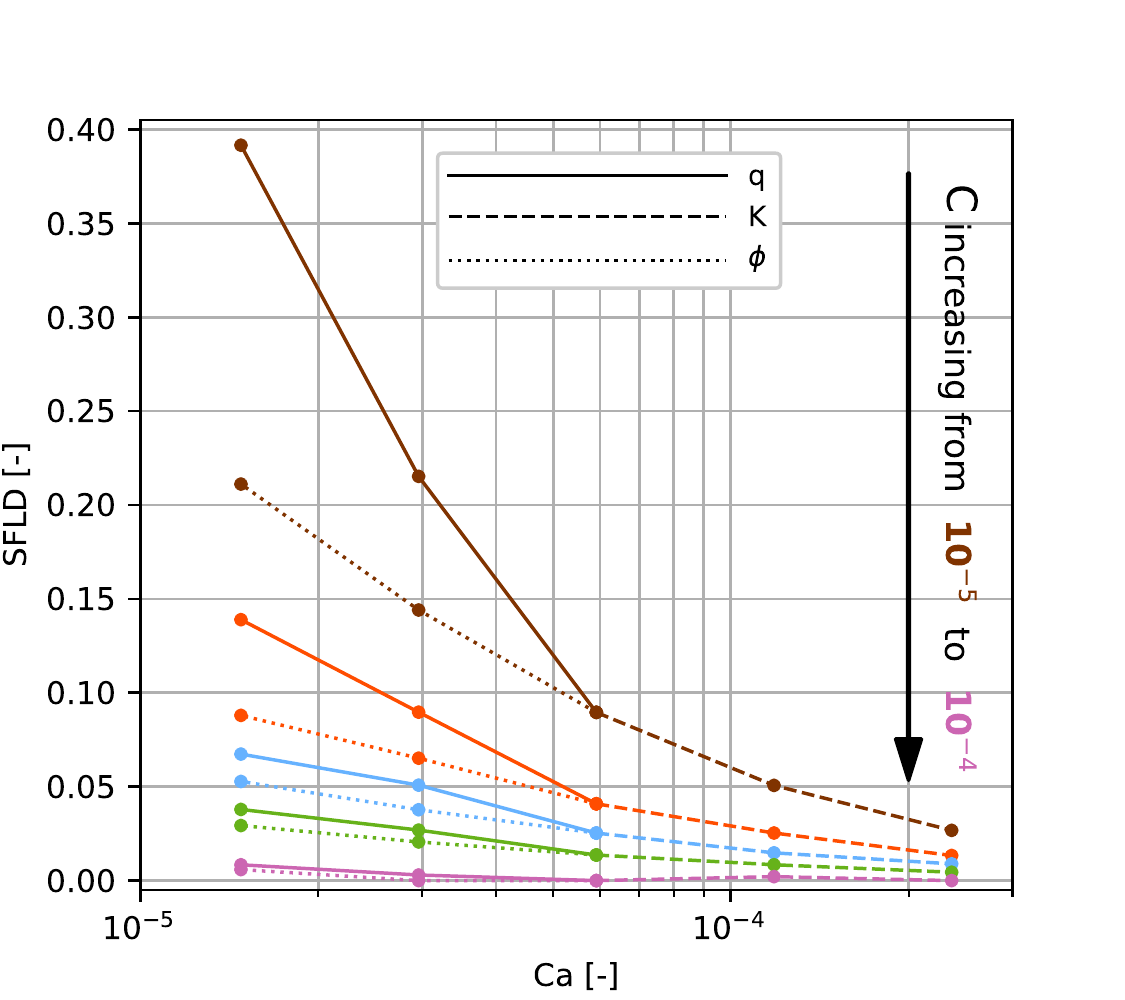}
\caption{1D-H scaled \co front location difference (SFLD) as a function of capillary number ($\mathcal{C}a$) for the combined effect on the dynamics in the saturation functions using the values in Table \ref{Param_1d_com}.}\label{dynamic1DRel2}
\end{figure} 

For impact of $\mathcal{C}a$ on the combined dynamics shown in Figure \ref{dynamic1DRel2}, we observe an increased impact at low capillary number when dynamics are modeled in both saturation functions. For example, the impact of WA dynamics in SFLD doubles for $C=2.5 \times 10^{-5}$ when relative permeability dynamics are added to capillary dynamics, with the increase being greater at higher capillary number. Thus, relative permeability dynamics help to compensate for the disappearing impact of dynamic $P_c$ at higher capillary number. {However, the impact of the relative permeability dynamics is very minor, which implies that using less extreme relative permeability curves more reflective of laboratory studies will have negligible impact on \co migration under dynamic wettability conditions. }

 Intuitively, the results in Figures \ref{SFLD_Ca} and \ref{dynamic1DRel2} reflect the fact that very viscous flows will not be greatly impacted by wettability dynamics in either saturation function. The minimal effect of capillarity at high capillary number is expected given that capillary effects disappear with higher $\mathcal{C}a$, and therefore any additional dynamics in $P_c$ have a negligible effect. But we also see that the impact of relative permeability dynamics also decreases with higher capillary number albeit more slowly. For \co  storage settings, these observations point to the increased importance of wettability dynamics farther from the injection well where capillary forces are likely to be dominant.
 
\subsection{1D-vertical (1D-V) flow system}\label{2dexamp}

In this section, we consider a 1D vertical cross-section of an aquifer-caprock system. This system is employed to demonstrate the impact of WA on containment of  \co beneath an initially sealing caprock. This system is modeled as a one-dimensional flow domain of $A= 10$ m$^2$ cross-sectional area by $H =100$ m height. The system is discretized into 100 elements along the height.
The top and bottom zones have contrasting permeability and capillary entry pressure. The permeability and porosity in the aquifer are homogeneous and fixed at 10$^{-10}$ m$^2$ and 0.2, respectively. The caprock permeability is homogeneous and varied in different simulations (see Table \ref{Paracom2d}), while porosity is fixed at 0.2.
The caprock is initially water-wet and altered by exposure to dissolved \co, such that the wettability altering agent is $X^{\mbox{WA}}=X^{\mbox{\co}}_w$ for all 1D-V simulations.

The saturation functions in the aquifer are set equal to the final state and are kept static. We employ the same end-wetting state parameters used in the horizontal study (refer to Table \ref{BtubeM_End}) for the aquifer. On the other hand, different initial-wet entry pressure and permeability values are tested for the caprock section (see Table \ref{Paracom2d}).

\begin{figure}[h!]
    \centering
\includegraphics[scale=0.41]{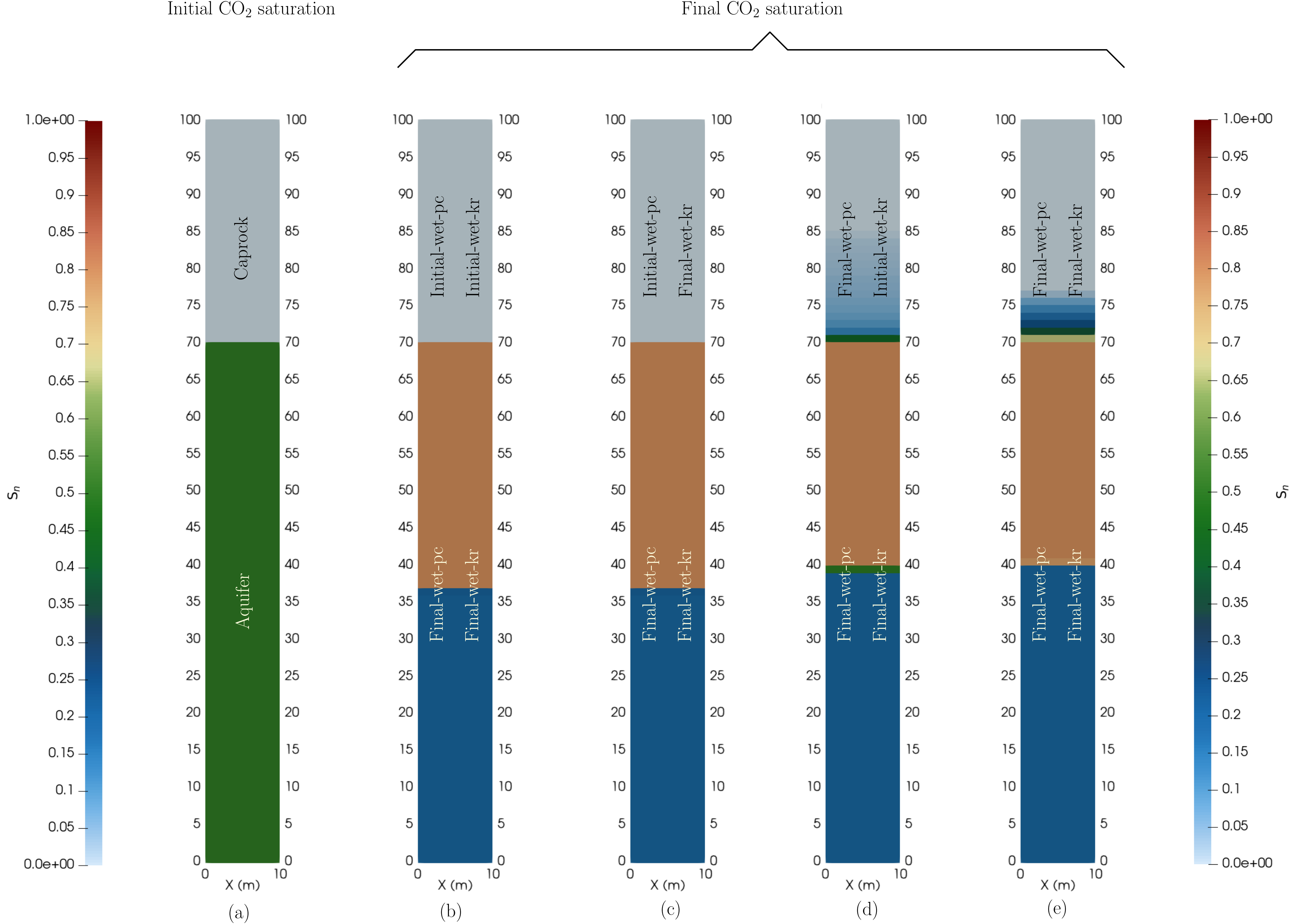}   
 \caption{1Dvfs (a) initial \co saturation and final \co saturation distributions for the (b) initial-pc-initial-kr functions, (c) initial-pc-final-kr functions, (d)  final-pc-initial-kr functions, and (e) final-pc-final-kr functions after 100 years.}
    \label{verticalvertical}
\end{figure}

The initial condition, depicted in Figure \ref{verticalvertical}a, is set such that a 50\% \co saturation is uniformly distributed in a column of height $h=70$ m. The reservoir pressure is considered to be hydrostatic with regard to brine density and depth of the formation. For this case, the temperature and salinity of the reservoir are considered to be 20\textdegree{}C and 0 ppm, respectively. The diffusion coefficient of \co in the brine wetting phase is set to $2\times 10^{-9}$ m$^2$/s. All boundaries are closed to flow, and the residual saturation of each phase is $S_{r\alpha} =0.2$. \co mole fraction in the brine phase is set initially to $5\times10^{-3}$.

The above described condition is initially not at equilibrium due to capillarity and gravity gradients. In the absence of any other driving forces, the \co and brine will redistribute in the aquifer according to buoyancy and capillarity to reach an equilibrium. Simultaneously, \co dissolves into the brine and
diffuses into the caprock, altering the capillary entry pressure over time. Since the capillary pressure in the aquifer is small, the equilibration in the aquifer with regard to gravity is in order of days, while the significant WA effects in the caprock starts in the order of years.

\noindent We begin with analysis of the static wettability system for different combinations of initial- and final-wetting states for each saturation function. Figures \ref{verticalvertical}b-e show the \co distribution along the column after 100 years for each combination. {The first two cases Figures \ref{verticalvertical}b and \ref{verticalvertical}c maintain water-wet conditions in the capillary pressure curve. It is clear that the capillary seal is sufficient to contain \co under the caprock, and containment is not sensitive to the parameters of the relative permeability functions. We also observe the resulting redistribution of \co in the aquifer from the initial condition in Figure \ref{verticalvertical}a results in a near complete gravity segregation of \co and brine, leaving residual \co (with saturation of 0.2) below the accumulated \co column. } 

{In comparison, when the end-wetting capillary pressure curve is directly applied in the caprock (Figures \ref{verticalvertical}d-e), the capillary seal is no longer sufficient to contain \co. The difference between Figures \ref{verticalvertical}d and \ref{verticalvertical}e shows the extent to which the wetting properties of the relative permeability affects \co migration in the caprock over time. This result shows if the wettability state of an initially water-wet caprock is changed by exposure to dissolved \co, then \co can eventually migrate into the caprock. The dynamics of the wettability change coupled with diffusion of dissolved \co upward into the caprock will ultimately determine the behavior of \co in this system over time. }

\begin{figure}[h!]
    \centering
    \includegraphics[scale=0.55]{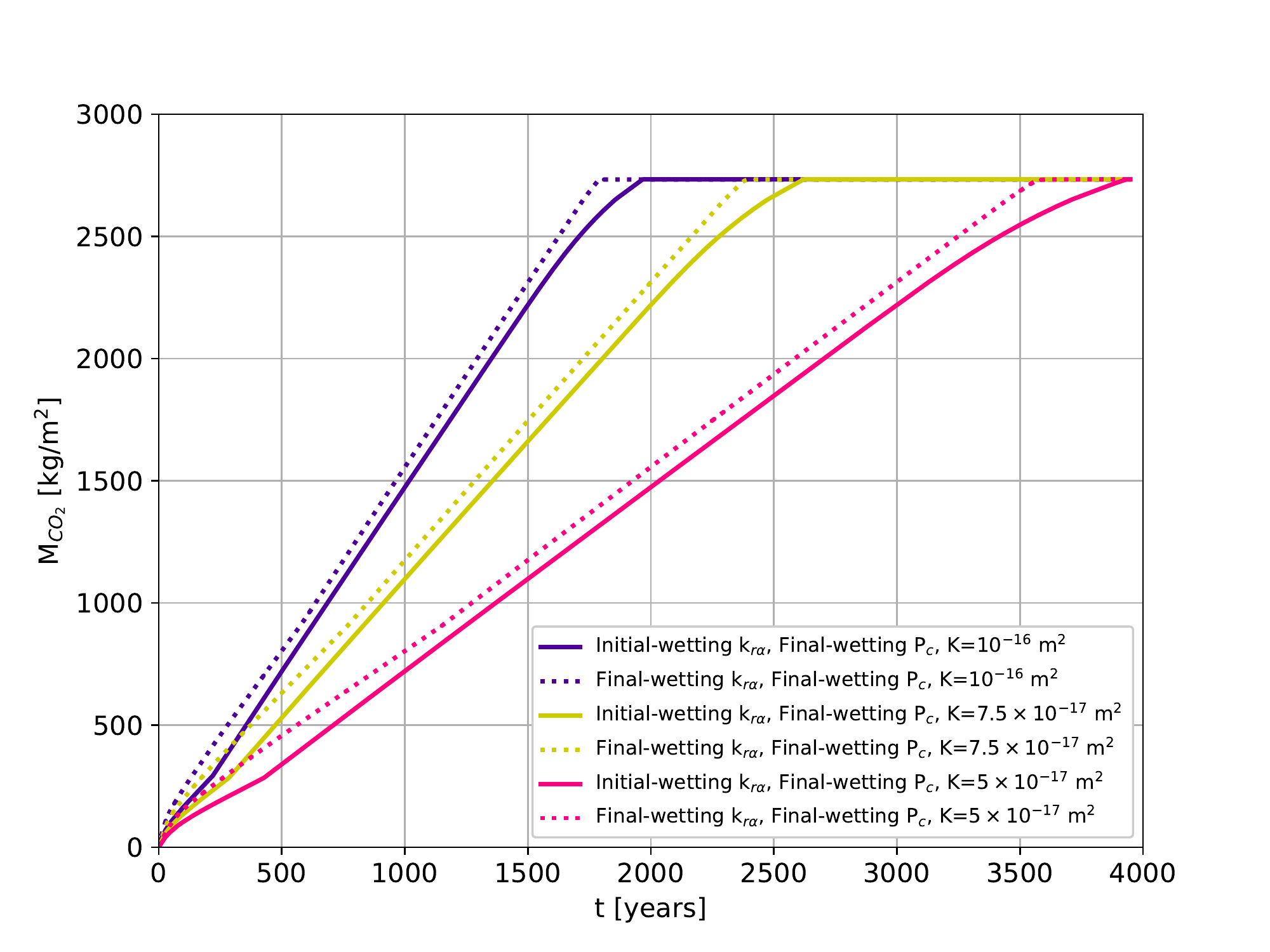}
    \caption{The evolution of \co mass over time in the caprock.}
    \label{mas_over_time}
\end{figure}

Given the dimensions and initial conditions of this closed system, then the maximum amount of \co migrating into the caprock can be estimated. Neglecting the dissolved \co in the brine and assuming the non-wetting phase is mostly \co, then the initial mass of \co in the system is $M_{\co}^{\rm i}/A\approx h\rho_n S_n^{\rm i}\phi=5\times 10^3$ kg/m$^2$. Considering the residual \co in the aquifer, this results in a maximum migration of \co into the caprock of ca. $M_{\co}^{\rm max}/A\approx 3\times 10^3$ kg/m$^2$. Figure \ref{mas_over_time} shows the accumulated \co in the caprock over time for the combinations of final capillary pressure function with the initial- and final-relative permeability functions for three different caprock permeability values. As expected, we observe that the \co migrates faster into the caprock for the final-wetting functions, the lower the rock permeability the slower the \co migrates into the caprock, and the amount of \co is limited by the initial and residual saturation ($M_{\co}^{\rm max}/A= 2733$ kg/m$^2$). 

\noindent {We now introduce dynamics into the wetting state of the caprock.} In this vertical system, the WA agent is modeled as dissolved mass fraction, i.e.,  $X^{\mbox{WA}}=X_w^{\mbox{\co}}$. We recall that the alteration process, or the measure of exposure time, depends on the magnitude of the WA agent. The dissolved \co mass fraction in water is of order 10$^{-2}$ while the non-wetting saturation is of order 10$^{-1}$.  To observe the WA effects within years in the 1D-V studies, we choose relatively higher values for model parameters $b_1$, $b_2$, $\nu_1$, and $\nu_2$ (Table \ref{Paracom2d}) in comparison to the ones we considered in the 1D-H simulations (Table \ref{BtubeM}).   

\begin{table}[h!]
\caption{Parameters describing the relationship between the pore-scale parameter $C$ associated with contact angle change and dynamic coefficients $\beta$ and $\eta$ for the 1Dvfs ($t_{ch}=10^7$ s, $X^{WA}=X_w^{\rm CO_2}$).}\label{Paracom2d}
\centering
\begin{tabular}{ c c c c c c c}
\hline
$b_1$   & $b_2$  & $\nu_1$ & $\nu_2$ & $C$  & $\beta=b_1C^{b_2}$ & $\eta=-\nu_1 C+\nu_2$      \\
$[-]$   & $[-]$  & $[-]$ & $[-]$   & $[-]$  & $[-]$  & $[-]$           \\
\hline 
&&&&$10^{-5}$ & $10^{-2}$ & 4.5$\times 10^{-1}$\\
$10^{7}$ &   1.8     &    $4.999\times 10^{3}$    &  $5\times10^{-1}$ &$2.5\times10^{-5}$& 5.2$\times 10^{-2}$ & 3.75$\times 10^{-1}$\\
&&&&5$\times10^{-5}$& 1.81$\times 10^{-1}$ & 2.5$\times 10^{-1}$\\
\hline
\end{tabular}
\end{table}

Similar to the studies in Section \ref{1dSimExa}, we perform a sensitivity analysis to quantify the impact of dynamic saturation functions on \co migration into the caprock by varying rock properties (permeability and initial entry pressure) and dynamic parameters ($\beta(C)$ and $\eta(C$). Here, we examine the impact of WA on the integrity of caprock by
\begin{itemize}
    \item Case 1:  considering WA dynamics only in the capillary pressure function and
    \item Case 2:  considering WA dynamics in both the capillary pressure and relative permeability functions.
\end{itemize}

Table \ref{Param_vertical} presents the parameters combinations for the quantification of the two cases mentioned above.

\begin{table}[h!]
\caption{Tested parameter values used in the 1D vertical simulation study. A set of simulations were performed for each individual value of $c^{\rm i}$ and $K$ in which $C$ is varied, where the isolated (only capillary pressure) and combined impact on dynamic wettability functions is studied. For each value of $C$, the corresponding value of $\beta$ and $\eta$ is given in Table \ref{Paracom2d}.  }\label{Param_vertical}
  \centering
    \begin{tabular}{c c c c}
        \hline
        $c^{\rm i}$ & $K$ & $C$  \\
        Pa &  m$^2$ &  $[-]$ \\
        \hline
        $10^{4}$ & $1 \times 10^{-16}$  & $10^{-5}$\\
        $5\times10^4$ & $7.5\times10^{-17}$  & $2.5\times10^{-5}$\\
        $10^5$ & $5\times10^{-17}$  & $5\times10^{-5}$\\
          \hline
    \end{tabular}
\end{table}

\begin{figure}[h!]
    \centering
    \subfigure[]{\includegraphics[scale=.49]{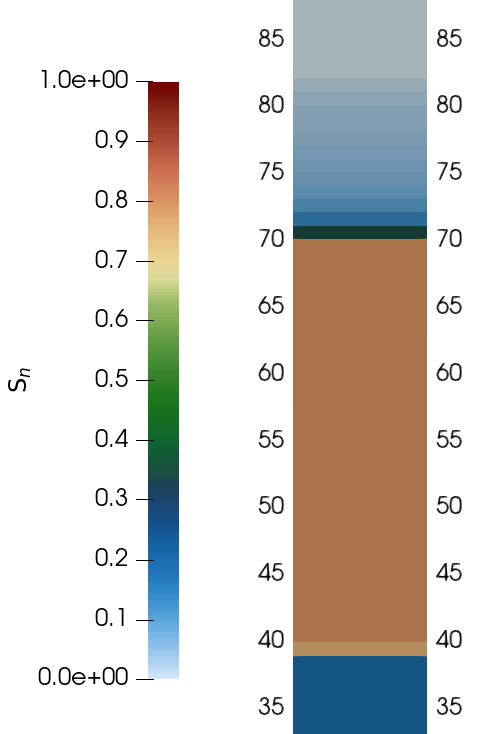}}
    \subfigure[]{\includegraphics[scale=.49]{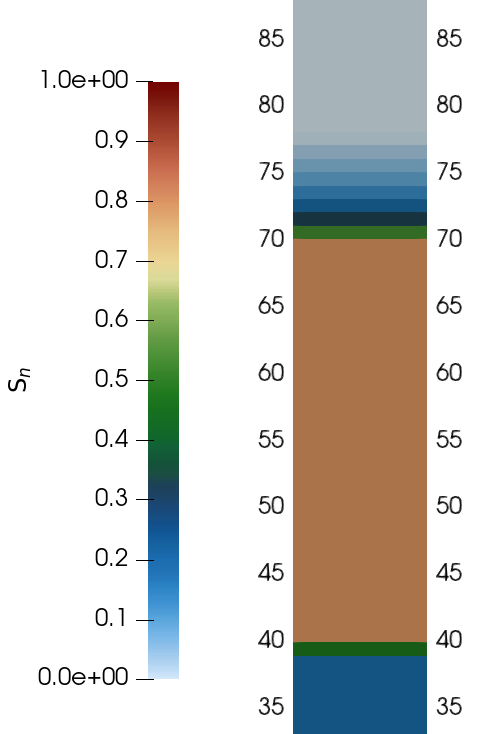}}
    \subfigure[]{\includegraphics[scale=.67]{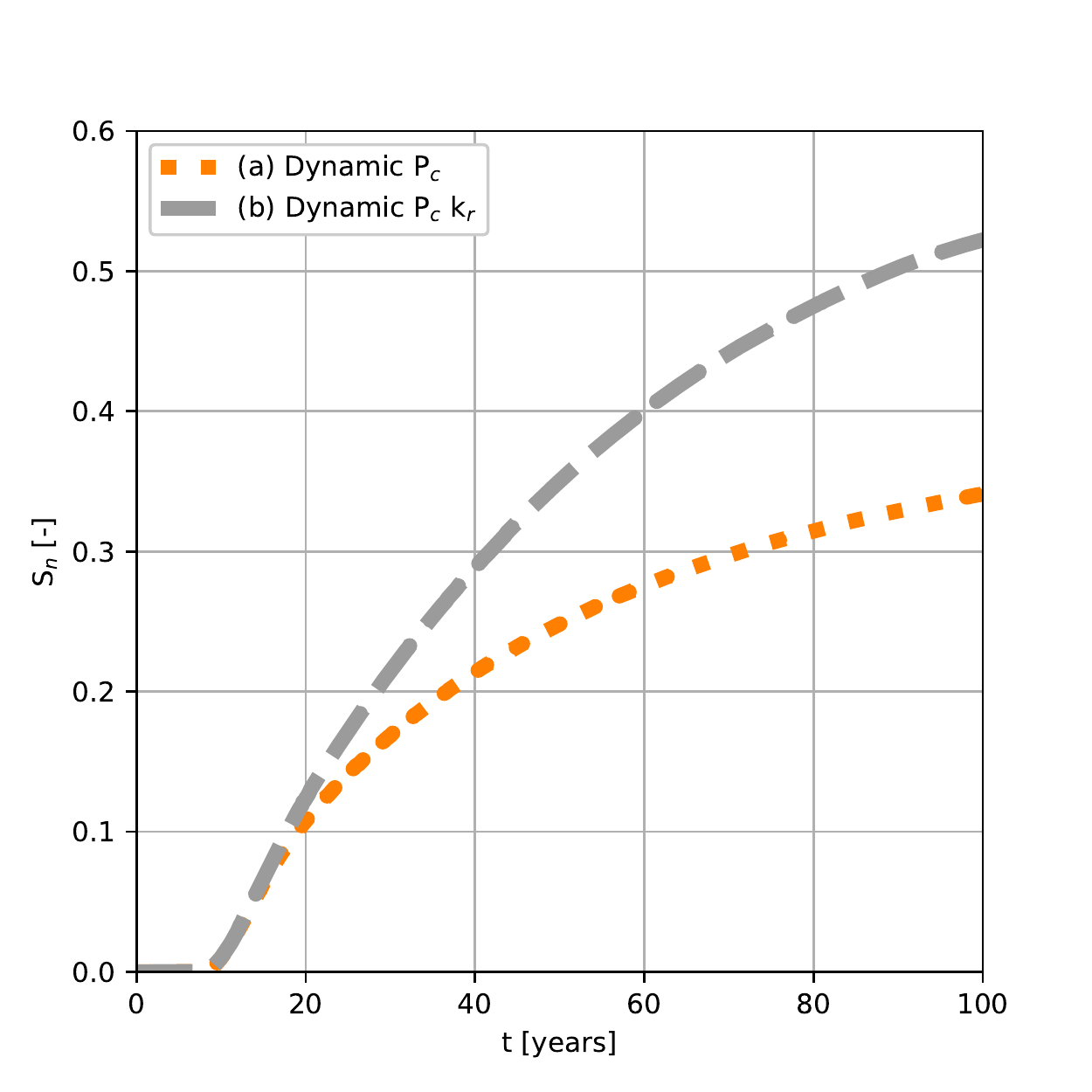}}
    \caption{(a) \co-water distribution in a column after 100 years given dynamic capillary pressure only and (b) dynamic capillary pressure and relative permeability functions. (c) The evolution of \co over time for the first grid cell on the caprock for cases (a) and (b).}
    \label{dynacbetadis}
\end{figure}

\noindent An example of the impact of dynamics in wettability on the vertical \co distribution is shown in Figure \ref{dynacbetadis} for the parameter combination of $c^i$=10$^4$ Pa, $K=10^{-16}$ m$^2$, and $C=5\times 10^{-5}$ under two different dynamic cases. For Case 1 (Figure \ref{dynacbetadis}a), the caprock capillary entry pressure is reduced by exposure to dissolved \co, allowing mobile \co to migrate upwards into the low permeability domain. For Case 2 (Figure \ref{dynacbetadis}b), an added effect occurs in the caprock. \co migrates upwards into the caprock at a higher saturation when relative permeability is altered, which is in agreement with the shift in fractional flow function for vertical flow from initial to final wetting condition (Figure \ref{fig:fracflow}d). Figure \ref{dynacbetadis}c shows the evolution of WA induced \co saturation over time in the first grid cell within the caprock domain. For this set of parameters, we observe breakthrough into the caprock after approximately 10 years. Once \co begins to migrate vertically, both cases are characterized by a steep increase followed by a more gradual accumulation of \co in the selected grid cell. 

\begin{figure}[h!]
    \centering
    \includegraphics[scale=0.4]{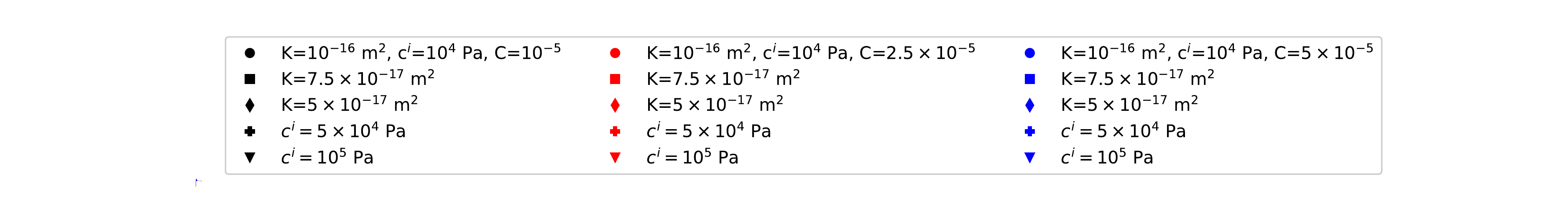}\\[-.5cm]
    \subfigure[]{\includegraphics[scale=0.41]{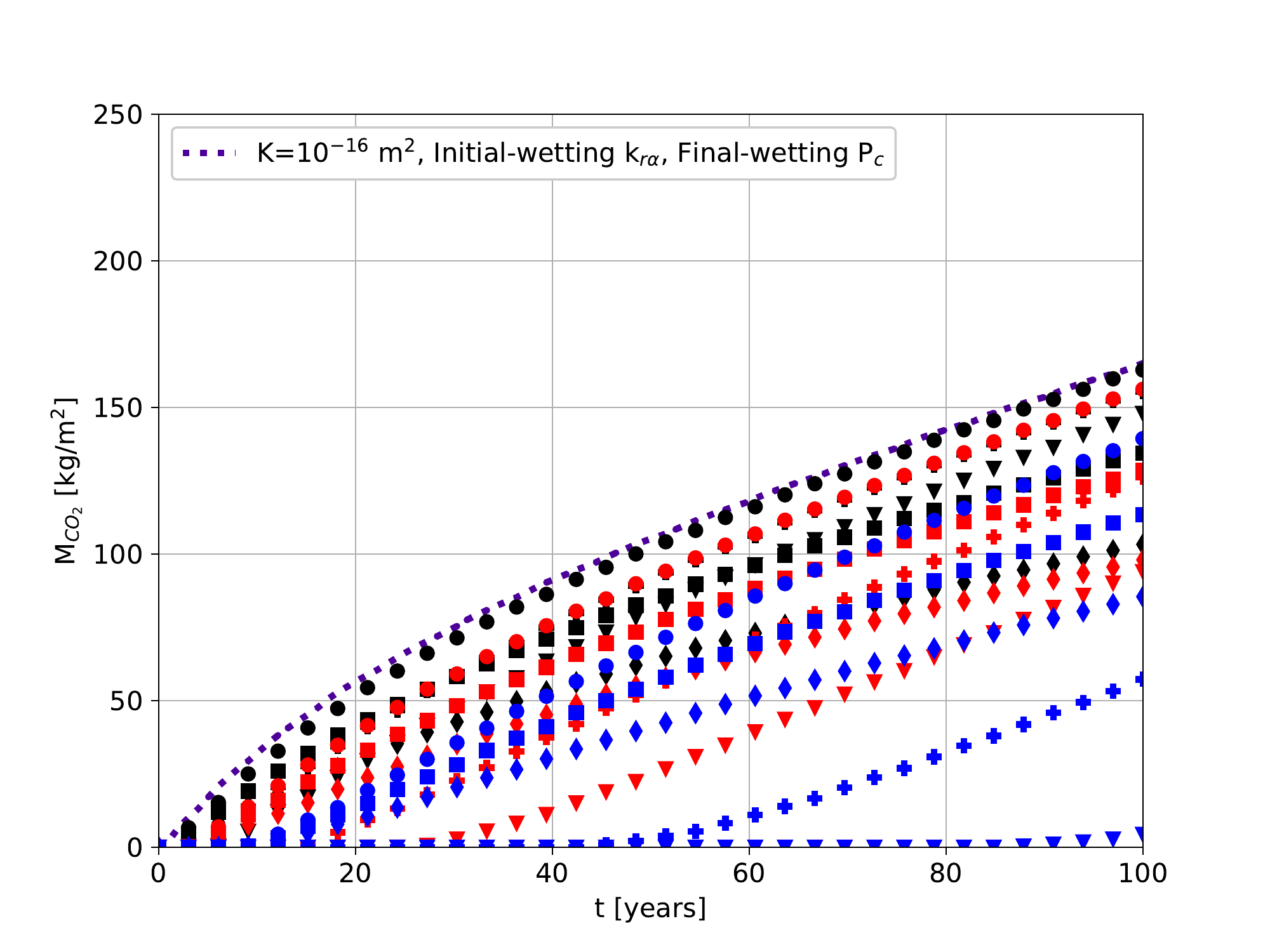}}
    \subfigure[]{\includegraphics[scale=.41]{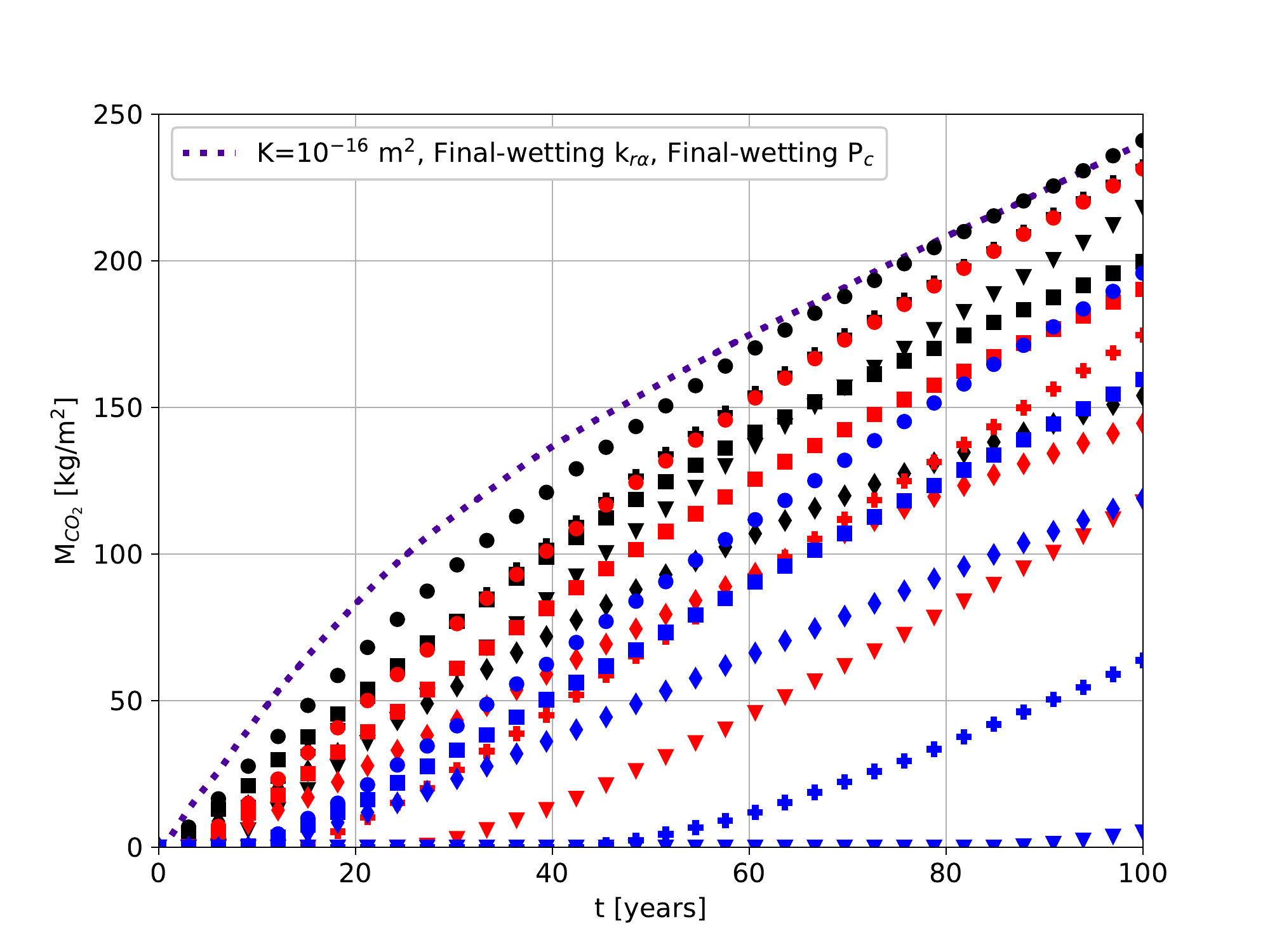}}
    \caption{Total \co mass in the caprock over time for each simulated parameter combination: (a) Case 1: dynamic $P_c$ and (b) dynamic $P_c$ and $k_{r\alpha}$.  }
    \label{SD}
\end{figure}

A set of simulations were performed for each individual value of $c^{\rm i}$ and $\rm K$ in which $C$ is varied (Table \ref{Param_vertical}). These numerical results are shown in Figure \ref{SD}a for Case 1 (isolated $P_c$ dynamics) and Figure \ref{SD}b for Case 2 (combined dynamics). Comparing the two cases, we observe the build-up of \co follows qualitatively the same response to changes in parameter values, but where the amount of \co in the caprock is less when dynamics are limited to capillary pressure function (Case 1, Figure \ref{SD}a) than when dynamics are presented in both saturation functions (Case 2, Figure \ref{SD}b). 

As expected, the larger the value of caprock entry pressure, the longer it takes for the \co to start entering the caprock. This starting migration time increases with smaller values of $C$ (slower dynamics on the saturation functions). On the other hand, smaller values of rock permeability result in slower \co migration into the caprock (see also Figure \ref{mas_over_time}), which is also affected by the value of the pore-scale parameter $C$ and entry pressure $c^{\rm i}$ as observed in Figure \ref{SD}.

Given the above observations, it may be possible to obtain a general relationship for the impact of WA on \co containment. We first divide the variables by reference values to make them dimensionless. The reference curves for the scaling are the ones giving by the static saturation functions (purple curves in Figures \ref{mas_over_time} and \ref{SD}). We have tested different functions for the scaling and selected the ones we describe next. For the time variable, we suggest a translation as a function of the entry pressure and pore-scale parameter: 
\begin{eqnarray}\label{eq:scale_t}
\hat{t}=\frac{t}{t_{\rm ref}}-a_0\frac{c^{\rm i}}{c_{\rm ref}}\left(\frac{C}{C_{\rm ref}}\right)^{a_1},
\end{eqnarray}
where a$_0$ and a$_1$ are fitting parameters. For the \co mass in the caprock, we suggest the following scaling function:
\begin{equation}
   \hat{M}_{\rm \co}=\frac{M_{\rm \co}}{M_{\rm \co}^{\rm max}}\left[\left(\frac{K}{K_{\rm ref}}\right)^{a_2}\left(\frac{C}{C_{\rm ref}}\right)^{a_3}+a_4\left(\frac{c^{\rm i}}{c_{\rm ref}}\right)^{a_5}\left(\frac{C}{C_{\rm ref}}\right)^{a_6}\right],
\end{equation}
where a$_2$,\ldots, a$_6$ are fitting parameters.

From Figure \ref{mas_over_time} we observe a linear behavior between \co mass and time, until reaching a maximum value of mass. This motivates the following relationship between scaled mass and time:
\begin{equation}\label{model_CO2}
  \hat{M}_{\rm \co}=  \min\left (\psi_0\hat t,1\right )
\end{equation}
where $\psi_0$ is a fitting parameter.

\begin{figure}[h!]
     \centering
     \subfigure[] {\includegraphics[scale=0.41]{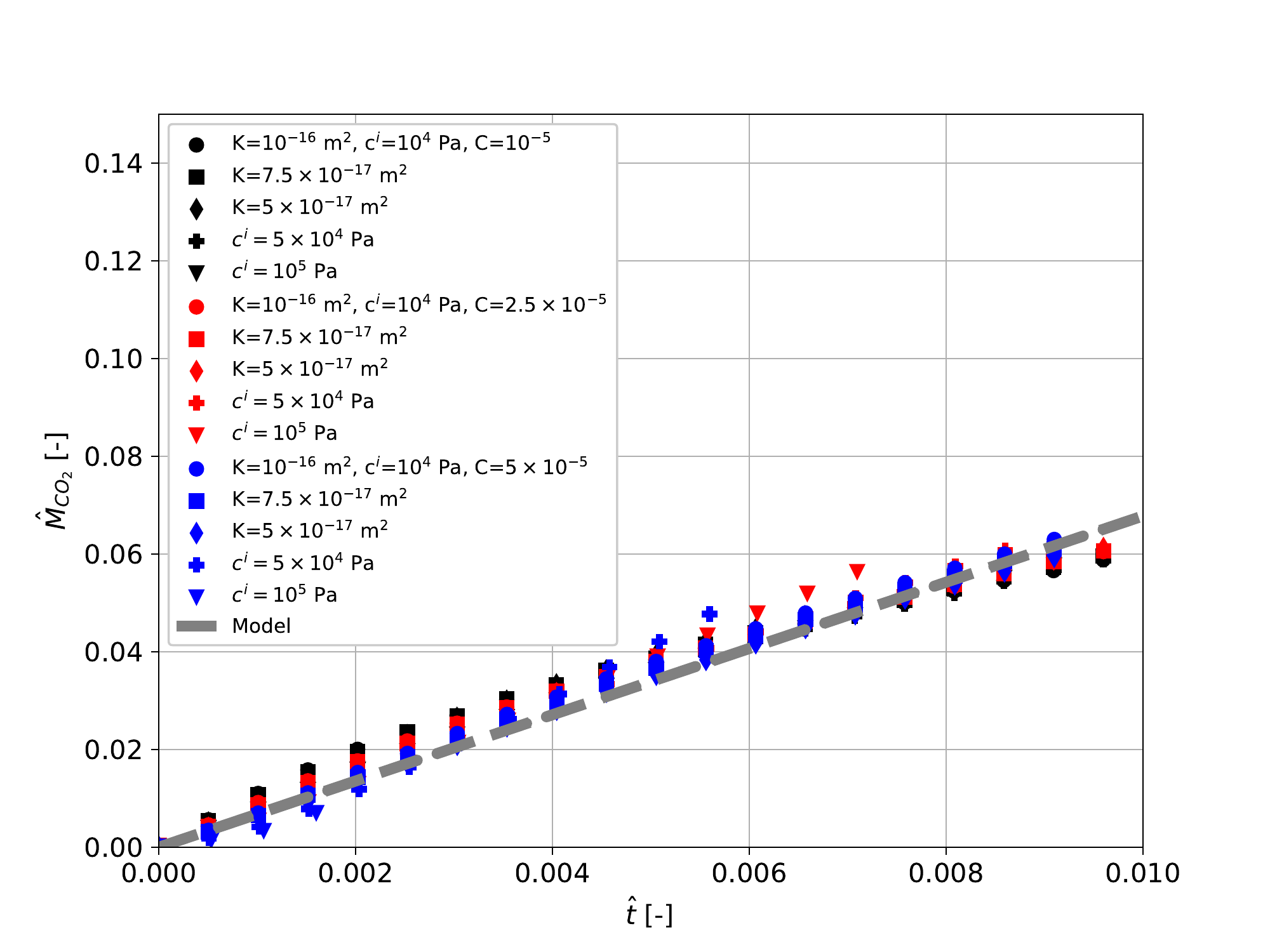}}
     \subfigure[]{\includegraphics[scale=.41]{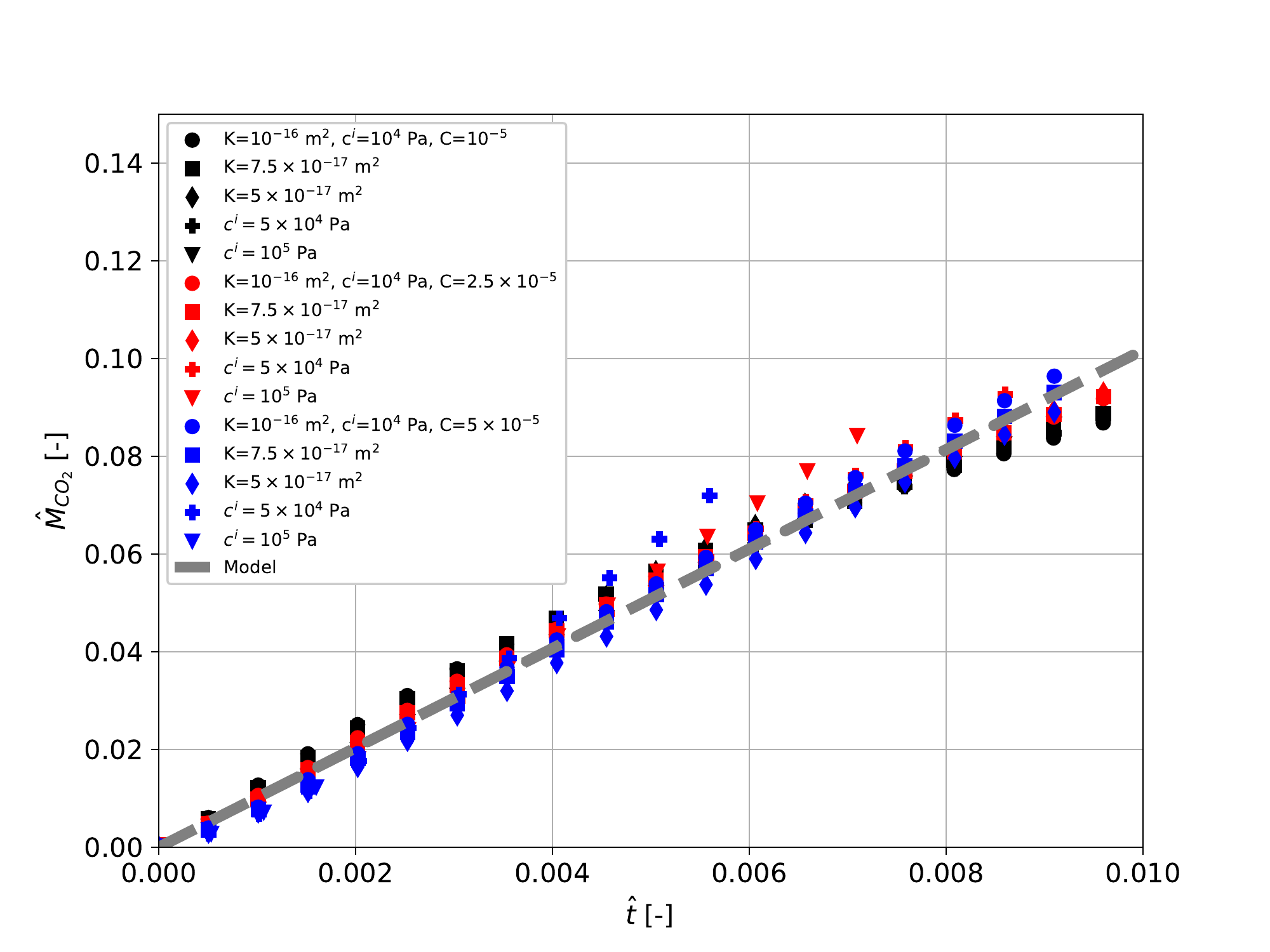}}
     \caption{The scaled amount of \co $\hat{M}_{\co}$ in the caprock along the scaled time $\hat{t}$ for the (a) dynamic $P_c$ and (b) dynamic $P_c$ and $k_{r\alpha}$.}
     \label{SD_scaled}
 \end{figure}

Table \ref{Fitting} shows the values of the reference and fitting parameters for both cases. Figure \ref{SD_scaled} shows the different scaled simulation results. Thus, we have obtained simple models to describe the impact of WA change on \co containment given caprock parameters $K$ and $c^{\rm i}$, and dynamic WA parameter $C$.

\begin{table}[h!]
\caption{Reference and fitting parameters.}\label{Fitting}
\centering
\begin{tabular}{l c c c c c c c c c c c c c c}
\hline
Case & $t_{{\rm ref}}$   & $c_{{\rm ref}}$  & $C_{{\rm ref}}$ & $K_{{\rm ref}}$ & $M_{\rm \co}^{{\rm max}}$  & $a_0$ & $a_1$ & $a_2$ & $a_3$ & $a_4$ & $a_5$ & $a_6$ & $\psi_0$  \\
&${\rm y}$   & ${\rm Pa}$  & $[-]$ & ${\rm m}^2$   & ${\rm kg/m}^2$  & $10^{-5}$  & $10^{0}$ &$10^{-1}$& $10^{-1}$& $10^{-3}$& $10^{0}$& $10^{0}$& $10^{0}$          \\
\hline 
1& \multirow{2}{*}{$10^{4}$} &   \multirow{2}{*}{$10^{4}$}  &    \multirow{2}{*}{$10^{-5}$}    &  \multirow{2}{*}{$10^{-16}$}&  \multirow{2}{*}{2733} & 5.3 & 1.7 & 4.9 & 1.2 & 2.0 & 1.6 & 2.4 &6.9\\
2& &    &    & & & 5.3 & 1.7 & 5.0 & 1.5 & 2.5 & 1.6 & 2.5 & 10  \\
\hline
\end{tabular}
\end{table}
 
\section{Discussion}

We have implemented dynamic capillary pressure and relative permeability functions to simulate and quantify the impact of WA on \co storage and containment. Horizontal and vertical field-scale test cases were used to demonstrate the effect of WA in \co front migration. These simplified scenarios allowed for testing of a wide range of different rock-fluid and wettability parameters in order to understand the impact of WA in relation to other multiphase flow processes.  Our findings show that long-term WA can be characterized according to capillary number for horizontal flow and  to caprock integrity parameters for vertical migration. These results can be used to understand the response of \co storage mechanisms to wettability alteration by long-term exposure to supercritical and dissolved \co.

\subsection{\co storage efficiency}
The horizontal case study that focused on the impact of WA on storage efficiency shows the importance of dynamics in both saturation functions, $P_c$ and $k_{r\alpha}$, during \co injection. Wettability change sharpens and delays the front movement, and thus a greater \co storage efficiency results with increasing exposure to injected \co. A defining feature of long-term WA is time-dependent heterogeneous wettability, resulting in a wettability gradient outwards from the injection well. Longer exposure time towards the inlet significantly alters wettability over time, while the leading edge of the \co front remains near water-wet. The interplay between dynamically changing  saturation functions and heterogeneous wettability leads to an increasingly complex evolution of \co migration with increased injected volumes, even for a simple 1D system. Altered wetting conditions can spontaneously draw \co back towards the inlet by capillary action. These interesting dynamics that may occur over months and years and can have important implications for storage efficiency in systems where WA is expected. More investigation is needed to fully understand the complex behavior caused by long-term WA in non-idealized storage systems.

The horizontal case study results also provide additional insight to the role of WA on storage efficiency as a function of capillary number. The advancing \co front, which can be considered a proxy for storage efficiency, can be significantly impacted at lower capillary numbers, but where the rate of dynamics in the saturation functions plays a important role. At higher capillary numbers, storage efficiency is much less impacted by long-term WA, regardless of the underlying parameters controlling the wettability dynamics. This result gives important insight into the flow regimes where long-term WA is a relevant process (and similarly, where WA dynamics could be neglected). \co storage systems often have a range of flow regimes across the domain at any given time, with highly viscous flows near the injection well and capillary-dominated flow further afield. 

{Another important finding is that the observed impact of wettability dynamics at the field-scale are controlled mainly by dynamics in the capillary pressure function and are mostly insensitive to the dynamics in relative permeability. We recall that the relative permeability dynamics we modeled represented a quite severe transition in curvature, and therefore the expected dynamics if using lab-derived functions will be even smaller. Therefore, it is reasonable to conclude that one can neglect dynamics in relative permeability and only focus on dynamics in capillary pressure for simulating field-scale impacts of long-term changes in wetting condition due to \co exposure. We emphasize that this conclusion only applies to dynamics in the curvature of the relative permeability functions, and the change in end-point residual saturation due to \co exposure is still a relevant aspect to incorporate at the field-scale. The possibility for including this process in future work is discussed more in Section 4.4. }

\subsection{\co containment}
The second vertical case with \co migration into the overburden due to long-term WA shows how the same dynamics in $P_c$ and $k_r$ impact storage containment. In contrast to the horizontal case, the vertical WA process is dependent on a certain sequence of effects: (1)  diffusion of dissolved \co into the caprock; (2) reduction  of caprock entry pressure by \co exposure that allows supercritical \co to break the capillary seal; (3) vertical migration of \co in caprock.  We note that the impact of altered relative permeability only acts once \co is mobile in the caprock. This sequence of processes implies that even if WA induces a loss of containment, it is first and foremost a slow process. \co continually encounters unexposed caprock along its vertical migration path that must be altered by the same slow process in order for vertical flow to continue. 

Although we observe that long-term WA by \co exposure leads to loss of containment, our simulations show that \co advances very slowly, but steadily, over time. Our scaling analysis shows that \co migration follows a characteristic evolution that is amenable to scaling by the underlying parameters, which gives insight into a unified model for long-term \co migration into a caprock due to long-term WA. The scaling model is a function of the rate of WA dynamics, caprock permeability, and initial capillary entry pressure. The model is an important generalization and valuable for making use of experimental/surveyed data to predict the integrity of caprocks exposed to \co over long timescales, i.e., one can quantify {\it a priori} the potential for unwanted \co migration without having to perform field-scale simulations. More importantly, this scaling relationship shows that long-term WA poses very little risk to \co containment. Only with a dramatic reduction in  the initial strength of the capillary seal and relatively high caprock permeability will \co migrate vertically in non-negligible quantities over several decades. For the Sleipner \co storage project where caprock permeability is on the order of 100 nanoDarcy and a plume that covers an area of 10 km$^2$, this would translate to ca. 5 tons over 200 years (assuming WA on both saturation functions, $c^{\rm i}$=3 MPa, and $C=10^{-5}$). 

\subsection{Numerical implementation of dynamic flow functions}
All numerical studies presented in this paper are conducted using the open-source framework OPM. One of the advantages of using OPM is that utilizes modern hardware architecture (i.e., multiple cores and vectorization). We extended the TCTP and flow simulators in OPM to include dynamics in the saturation functions (P$_c$ and k$_{r\alpha}$). Studies on the 1D-H system for large $\mathcal{C}a$ by increasing the injection flow rate would require a large domain to avoid the saturation front to reach the boundary, while keeping the grid size small to reduce numerical errors in the computation of the scaled front location difference (SFLD). Smaller values of $K$ also result in larger $\mathcal{C}a$ while keeping the saturation front closer to the injection well. However, using small grids close to the injection well result in convergence issues. This is expected since the general purpose of OPM Flow is for large field-scale simulations. Then, one could modify further the code (specially the well models) to conduct studies with smaller grids. The time scale for the numerical studies on the 1D-V system is of order of decades, which is computationally expensive for simulations using full TPTC models since dynamics in the saturation functions in addition to computations of phase compositions restrict the size of the time step. Further investigation on solution strategies for this system is required to reduce the computational cost (e.g., by not updating $\bar{\chi}$ after each time step but at certain times).

\subsection{Extension to realistic systems}

{This study has applied the dynamic WA model to relatively simple 1D systems under controlled conditions and homogeneous media. Additional effects become relevant if we are to extend this understanding to realistic 2D and 3D storage reservoirs. For example, \co migration in the reservoir is more complex when affected by gravity and reservoir heterogeneity. In the horizontal 1D systems, we observed a longer \co exposure at the inlet versus further out in the reservoir. With the introduction of gravity, this dynamic could lead to slower upward migration for \co injected down-dip or at the bottom of a thick reservoir. Heterogeneity either in permeability, porosity, or initial wettability will add complexity to dynamics in wettability that needs to be further studied. We note that differences expected in realistic systems apply mainly for migration within the reservoir itself. Vertical migration in a low-permeability caprock is primarily a 1D process even in 3D systems. The main difference in moving to 2D/3D is the possibility for varying \co column height under the caprock interface, which plays a role in the buoyancy forces acting on the rising \co.
}

{Another important aspect is the impact of wetting dynamics on \co trapping efficiency as a residual phase. We have not  accounted for dynamics in residual saturations in our simulations as the examples used in this study are mainly focused on drainage processes. However, real systems will experience imbibition, especially for \co storage dependent on migration-assisted trapping. Intermediate-wetting conditions will reduce the capillary trapping capacity of reservoir rocks, which leads to a greater portion of \co remaining in a free phase. The change in residual trapping from water- to \co-wetting conditions by \co exposure will be strongly coupled to the impacts observed due to dynamic alteration of the flow functions studied here. Although residual trapping has been characterized under static wettability, more work is needed to develop models for dynamic changes in residual trapping due to \co exposure and couple them to the dynamic flow functions implemented in this study. Future studies may address this by adapting the approach proposed by \cite{Kassa2019, Kassa2020}; however, with additional complexity to the porous medium (i.e., pore-network model instead of parallel capillary tubes). 
}

\subsection{Additional pore-scale impacts}
{In this study, we have seen a marked difference in horizontal and vertical \co migration depending on the rate of wettability alteration as controlled by the pore-scale parameter $C$. We recall that the implemented dynamic flow functions have been upscaled from pore to core scale. Our results show how pore-scale dynamics are manifested at the field scale and the importance of quantifying $C$ through laboratory experiments. To date, only a few experiments have measured contact angle change through exposure to \co over long timescales, but none have described the temporal evolution of contact angle needed to calibrate pore-scale CA change models such as Equation \eqref{CA_model}. More experimental evidence is needed to determine if continual \co exposure could fully transform a water-wet rock to strongly \co-wet over the long timescales relevant for \co storage.
} 

{There are other pore-scale effects that can be of interest for further study, particularly with regard to geochemical impacts. First, \co-induced reactions can dissolve and precipitate minerals, thus altering the pore topology of reservoir rocks and caprocks \citep{Landa-Marban2021}. Changes in pore topology could alter the flow functions in a similar manner as changes in contact angle, but characterization of the connection between topological alteration and capillary pressure/relative permeability curves requires further investigation. Geochemical alteration can also lead to an increase in permeability, which could enhance vertical \co migration in addition to WA. On the other hand, an increased porosity can counteract the permeability increase. Further study is needed to quantify and characterize the combined impact of geochemical and wettability alteration on \co migration. }

\section{Conclusion}
In this work, we present a macroscale TPTC mathematical model to study long-term WA effects in CO$_2$ storage applications. Particularly, this model includes macroscale dynamic capillary pressure and relative permeability functions derived from pore-scale WA models. We use OPM to implement the model and perform the numerical studies. Two simple 1-D systems are considered to investigate the WA effects on different flow regimes.        

Based on this work our conclusions are as follows:
\begin{itemize}
    \item Horizontal and vertical field-scale test cases can be used to demonstrate the effect of long-term WA in CO$_2$ front migration.
    \item Long-term WA can be characterized according to capillary number for horizontal flow and to caprock integrity parameters for vertical migration.
    \item {Field-scale impacts of long-term WA are mostly controlled by dynamics in the capillary pressure function due to \co exposure, while similar dynamics in the curvature of the relative permeability functions is a secondary factor.}
    \item Scaling models to quantify CO$_2$ migration into the caprock show that long-term WA poses little risk to CO$_2$ containment.
\end{itemize}
 
\section*{Notation}
\vspace{-.2cm}
\begin{longtable}{l l}
{$\phi$}&{Porosity}\\
{$b_1$, $b_2$}&{Fitting parameters for the dynamic parameter $\beta$}\\
{$\nu_1$, $\nu_2$}&{Fitting parameters for the dynamic parameter $\eta$}\\
{$C$}&{Pore-scale parameter}\\
{$\rho_w, ~\rho_{n}$}&{Densities (wetting and non-wetting phase)}\\
{$S_w,~S_{n}$}&{Saturations (wetting and non-wetting phase)}\\
{$X_w^{k}, ~X_{n}^{k}$}&{Component $k$ mass fraction (wetting and non-wetting phase)}\\
{$\vec{u}_w,~\vec{u}_{n}$}&{Darcy flux (wetting and non-wetting phase)}\\
{$F^k$}&{Component $k$ source term}\\
{$\mathbb{K}, ~{K}$}&{Intrinsic permeability (tensor and scalar)}\\
{$\vec{j}_w^k,~\vec{j}_{n}^k$}&{Component $k$ diffusive flux (wetting and non-wetting phase)}\\
{$k_{rw}, ~ k_{rn}$}&{Relative permeability functions (wetting and non-wetting phase)}\\
{$\vec{g}$}&{Gravitational constant}\\
{$P_{w},P_{n}$}&{Pressures (wetting and non-wetting phase)}\\
{$f_n$}&{Non-wetting fractional flow function}\\
{$D^k_w$, $D^k_n$}&{Component $k$ molecular diffusion parameter (wetting and non-wetting phase)}\\
{$T$}&{Reservoir temperature}\\
{$\tau$}&{Tortuosity}\\
{$\mathbb{R}$}&{Set of real numbers}\\
{$t_{ch}$}&{Characteristic time}\\
{$f_w^k,~f_{n}^k$}&{Component $k$ fugacities (wetting and non-wetting phase}\\
{$\beta, ~\eta$}&{Dynamic parameters (capillary pressure and relative permeability functions)}\\
{$\lambda, ~\Lambda$}&{Fitting parameters (capillary pressure and relative permeability functions)}\\
{$c^{\rm i},~c^{\rm f}$}&{Entry pressures (initial and final)}\\
{$\theta^{\rm i},~\theta^{\rm f}$}&{Contact angle (initial and final)}\\
{$x^{\rm i},~x^{\rm dy}$}&{Non-wetting phase front location (initial and dynamical saturation functions)}\\
{$P_c^{\rm i},~P_c^{\rm f}$}&{Capillary pressure functions (initial and final)}\\
{$S_{wr}, ~S_{nr}$}&{Residual saturations (wetting and non-wetting phase)}\\
{$S_{we}$}&{Wetting phase effective saturation}\\
{$\chi,~\overline{\chi}$}&{Cumulative measure of exposure time (pore and Darcy scale)}\\
{$E^{\rm i}$, $E^{\rm f}$}&{Wetting-state parameters for the relative permeability functions (initial and final)}\\
{$\mu_w,~\mu_{n}$}&{Viscosities (wetting and non-wetting phase)}\\
{$\Omega$}&{Spatial domain (reservoir)}\\
{$\lambda_{w},~\lambda_{n}$}&{Mobilities (wetting and non-wetting phase)}\\
{$\vartheta$}&{Flow inclination angle}\\
{$\mathcal{C}a$}&{Capillary number}\\
{$\Delta x$}&{Length of a grid cell}\\
{$A$}&{Cross-sectional area (reservoir)}\\
{$q$}&{Injection rate}\\
{$h,~H$}&{Height (aquifer and aquifer+caprock)}\\
{$L$}&{Length (aquifer)}\\
{$M_{\rm CO_2}^{\rm i}, ~M_{\rm CO_2}^{\rm max}$}&{CO$_2$ mass (initial and maximum)}\\
{$\mathbb{F}$}&{Dynamic function for the relative permeability functions}\\
\end{longtable}

\noindent\textbf{Data availability:} The numerical simulator OPM used in this study can be obtained at \url{https://github.com/OPM}. Link to complete codes for the numerical studies can be found in \url{https://github.com/daavid00/kassa-et-al-2021}.\vspace{.5cm}

\noindent\textbf{Acknowledgements:} This work was supported by the Research Council of Norway [grant number 255510] and CLIMIT-Demo/Gassnova [grant number 620073].

\bibliographystyle{plainnat} 
\bibliography{Reference}
\end{document}